\documentclass[a4paper,fleqn,usenatbib]{mnras}

\usepackage[T1]{fontenc}
\usepackage{ae,aecompl}

\usepackage{newtxtext,newtxmath}

\usepackage{hyperref}
\usepackage{graphicx}	       
\usepackage{xcolor}            

\def\bea{\begin{eqnarray}}
\def\eea{\end{eqnarray}}
\def\be{\begin{equation}}
\def\ee{\end{equation}}
\def\text#1{\mathrm{#1}}
\def\ms{M_\odot}
\def\mmax{M_\text{max}}
\def\lgi{L_\gamma^\infty}
\def\nbp{n_B^p}
\def\nbn{n_B^n}
\def\nbdu{n_B^{\rm DU}}
\def\mdu{M_{\rm DU}}
\def\mp{M^p}
\def\mn{M^n}

\title[Thermal states of neutron stars]
{Thermal states of neutron stars with a consistent model of interior}

\author[M. Fortin et al.]
{M. Fortin,$^1$\thanks{E-mail: fortin@camk.edu.pl}
G. Taranto,$^2$
G. F. Burgio,$^2$
P. Haensel,$^1$
H.-J. Schulze$^2$
and J. L. Zdunik$^1$\\
$^1$ N. Copernicus Astronomical Center, Polish Academy of Sciences,
Bartycka 18, 00-716 Warszawa, Poland
\\
$^2$ INFN Sezione di Catania, Dipartimento di Fisica, Universit\`a di Catania,
Via Santa Sofia 64, 95123 Catania, Italy
}

\date{Accepted XXX. Received YYY; in original form ZZZ}

\pubyear{2017}

\begin{document}
\label{firstpage}
\pagerange{\pageref{firstpage}--\pageref{lastpage}}
\maketitle

\begin{abstract}
We model the thermal states of both isolated neutron stars and accreting
neutron stars in X-ray transients in quiescence
and confront them with observations.
We use an equation of state and superfluid baryon gaps,
which are consistently calculated.
We conclude that the direct Urca process is required to be consistent
with low-luminous accreting neutron stars.
In addition, proton superfluidity and sufficiently weak neutron superfluidity
are necessary to explain the cooling of middle-aged neutron stars
and to obtain a realistic distribution of neutron star masses.
\end{abstract}



\section{Introduction}

The structure of a neutron star (NS) is determined by the nuclear forces
via the equation of state (EOS),
which has to be sufficiently stiff in order to support the
observed massive $2.0\,\ms$ pulsars \citep{Antoniadis2013,Fonseca}.
Nuclear forces determine also the composition of NS cores,
as well as the effective masses and superfluid gaps of baryons,
and are therefore crucial for the heat capacity and neutrino emission rate
in NS cores.
These two quantities govern the cooling of middle-aged ($\lesssim 10^5\,$yr)
isolated NS (INS)
and are also involved in the determination of the surface photon luminosity of
accreting NS in X-ray transients in quiescence (qXRT) \citep{YP04}.

Theoretical calculations of these basic properties of NS matter are
difficult and suffer from big uncertainties,
stemming from the deficiencies and approximations
in the many-body theory of baryonic matter,
and the lack of knowledge of strong interactions at high density
(many-body forces, effect of the quark structure of baryons, etc., see,
e.g., \citealt{NSbook}).
It seems thus reasonable to consider first the simplest
(minimal) model of NS with nucleon cores,
consisting of highly asymmetric nuclear matter permeated by a
degenerate quasi-free Fermi gas of electrons,
and above the muon threshold, also muons.
In the minimal model,
nucleons are treated as point-like,
and interact via a two-body potential and three-body forces.
The solution to the many-body problem is to be consistent with
laboratory nuclear data,
referring mainly to weakly asymmetric nuclear matter
and density close to normal nuclear density $n_0=0.16\,{\rm fm^{-3}}$.
Even for this model, however,
application to NS cores involves a huge extrapolation to $(6-8)\,n_0$,
at very large neutron excess.

Fortunately, high-density NS models can be confronted with astrophysical data:
measurements of NS masses and radii \citep{OF16,masses,Exorot},
surface temperature for INS of known or estimated age,
and thermal photon luminosity of qXRT
with an estimated time-averaged accretion rate onto the NS
(see \citealt{BY15a} for recent observational data).
This can hopefully help in unveiling the real physics of the NS core
and thus testing microscopic calculations starting from nuclear forces.

Recently, initial stages of such a program were carried out in
\cite{Baldo2014,Sharma2015,Taranto2016}.
The starting point was the AV18 two-body potential plus an Urbana model
of three-body interaction,
and the many-body theory was the Brueckner-Hartree-Fock (BHF) approximation.
The three-body interaction in nuclear matter was adjusted to nuclear saturation
data, and required to be consistent with the phenomenology of heavy ion
collisions and basic NS data \citep{Taranto2013}.
The model has at the saturation point a symmetry energy of 32 MeV
and a slope of 53 MeV,
which is in agreement with current nuclear constraints,
see, e.g., Figure 13 in \cite{Fortin2016}.

Then a unified EOS valid for the outer crust, inner crust,
and the core was calculated \citep{Sharma2015},
nucleon effective masses were obtained,
and neutrino emissivities
(assuming normal nucleons) calculated \citep{Baldo2014}.
Finally, superfluid gaps for neutrons and protons for
the same nucleon interaction were obtained and used to model the cooling
Cas A NS and the thermal states of middle-aged, $\sim (10^2-10^6)\,$yr,
INS \citep{Taranto2016}.

An important feature of the latter reference was the fact
that the BHF EOS allows for
the very efficient direct Urca (DU) cooling process to set in at a fairly low
baryon density $n_B=0.44\,\text{fm}^{-3}$ (proton fraction $x_p=0.136$),
corresponding to a NS mass of $1.1\,\ms$ \citep{Taranto2016},
in contrast to most other comparable studies in the literature
\citep{2005Gusa,2011PRLPage,2011MNRASYA,2011MNRASSHTE,2012Bla,MMa,Fortin2016},
which employ smaller nuclear symmetry energies
and later onset of the DU process.
Non-millisecond pulsars,
which were not recycled and thus have a mass near their birth mass,
have masses as low as $\sim (1.1-1.2)\,\ms$ \citep{OF16,masses}.
which is close to the theoretical minimum NS mass due to
thermal evolution \citep{YP04,Page09,Burgio}.
Thus with the EOS used in this work,
the DU process is operating in all currently observed NS.

Nevertheless in \cite{Taranto2016} a very satisfactory reproduction of all
current NS cooling data could be obtained for a certain choice of pairing
gaps.

In the present paper we extend the theoretical study of cooling middle-aged
INS initiated in \cite{Taranto2016} to the case of qXRT,
old NS of age $\sim (10^8-10^9)\,$yr that accrete matter
from their companion star during short periods of time resulting in
the heating of their interior.
Theoretical models for INS are represented by {\em cooling curves}
relating the effective surface temperature to the NS age.
A cooling curve corresponds to an assumed NS mass and composition of the
heat-blanketing envelope.
For qXRT sources theory yields instead {\em heating curves}
showing photon luminosity versus time-averaged accretion rate \citep{YP04}.
A heating curve is obtained for a given NS mass,
assumed composition of heat blanketing envelope,
and value of the total deep crustal heating per one accreted nucleon.

The article is organized as follows.
The microscopic model of NS interior is presented in Sect.~2.
The approach to the modelling of the thermal evolution and
of the stationary states of transiently accreting NS is described in Sect.~3.
Theoretical predictions for INS and qXRT are presented in Sect.~4.
Comparison with measurements of surface thermal luminosity
for middle-aged INS and of qXRT
are used to test the validity of the microscopic model
and to select an optimal model for the superfluid properties in the
NS core consistent with all the available observational data.
Sect.~5 summarizes our results and draws the conclusions.

\section{Microscopic model}

\subsection{EOS and composition of the core}

We calculate the EOS of nuclear matter within the BHF theoretical
approach \citep{BaldoBook}.
The starting point is the Brueckner-Bethe-Goldstone (BBG) equation
for the in-medium $G$-matrix,
whose only input is the nucleon-nucleon (NN) bare potential $V$,
\be
 G[n_B;\omega] =
 V + \sum_{k_a k_b} V {{|k_a k_b\rangle Q \langle k_a k_b|}
 \over {\omega - e(k_a) - e(k_b)}} G[n_B;\omega] \:,
\label{e:g}
\ee
where
$n_B=\sum_{k<k_F}$ is the nucleon number density,
$\omega$ is the starting energy,
and the Pauli operator $Q$ determines the propagation of
intermediate baryon pairs.
The single-particle (s.p.) energy reads
\be
 e(k) = e(k;n_B) = {k^2\over 2m} + U(k;n_B) \:,
\label{e:en}
\ee
where the s.p.~potential $U(k;n_B)$ is calculated in the so-called
{\em continuous choice} and is given by
\be
 U(k;n_B) = {\rm Re} \sum_{k'<k_F}
 \big\langle k k'\big| G[n_B; e(k)+e(k')] \big| k k'\big\rangle_a \:,
\ee
where the subscript $a$ indicates antisymmetrization of the matrix element.
Finally the energy per nucleon is expressed by
\be
 {E \over A} =
 {3\over5}{k_F^2\over 2m} +
 {1\over{2n_B}} \sum_{k<k_F} U(k;n_B) \:.
\ee

In this scheme, we use the Argonne $V_{18}$ potential \citep{v18}
as bare NN interaction $V$ in the BBG equation~(\ref{e:g}),
supplemented by a suitable three-nucleon force (TBF) introduced in
order to reproduce correctly the nuclear matter saturation point.
In this work we adopt the phenomenological Urbana-type UIX TBF
\citep{TBF1,TBF2,TBF3,microtbf} as input,
which consists of an attractive term due to two-pion exchange
with excitation of an intermediate $\Delta$ resonance,
and a repulsive phenomenological central term.

Further important ingredients in the cooling simulations are the
neutron and proton effective masses,
\be
 \frac{m^*(k)}{m} = \frac{k}{m} \left[ \frac{d e(k)}{dk} \right]^{-1} \:,
\ee
which we derive consistently from the BHF s.p.~energy $e(k)$, Eq.~(\ref{e:en}),
see Ref.~\citep{Baldo2014} for the numerical parametrizations.
Though their effect is not large compared to other uncertainties regarding
the cooling,
we like to stress that in this paper a consistent treatment is used.

\begin{figure}
\vspace{-6mm}
\centerline{\includegraphics[scale=0.8]{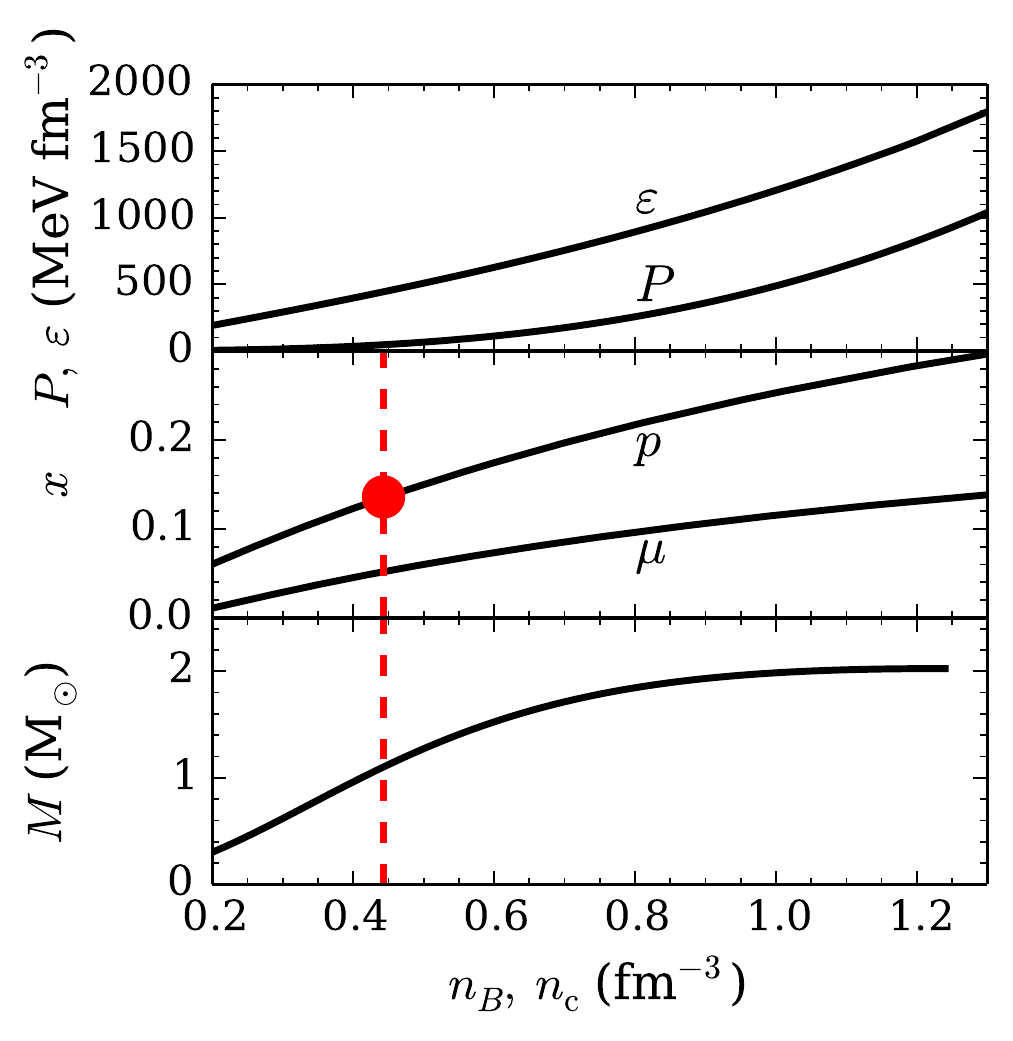}}
\vspace{-5mm}
\caption{(Color online)
Pressure and energy density (a),
and proton and muon fractions (b) are plotted
as functions of baryon number density $n_B$
in beta-stable matter for the BHF EOS.
In the lower panel (c) the neutron star mass is shown
as functions of the central density $n_c$.
The DU onset is indicated by the red dotted line.}
\label{f:eos}
\end{figure}

\begin{figure}
\vspace{-1mm}
\centerline{\includegraphics[scale=0.45]{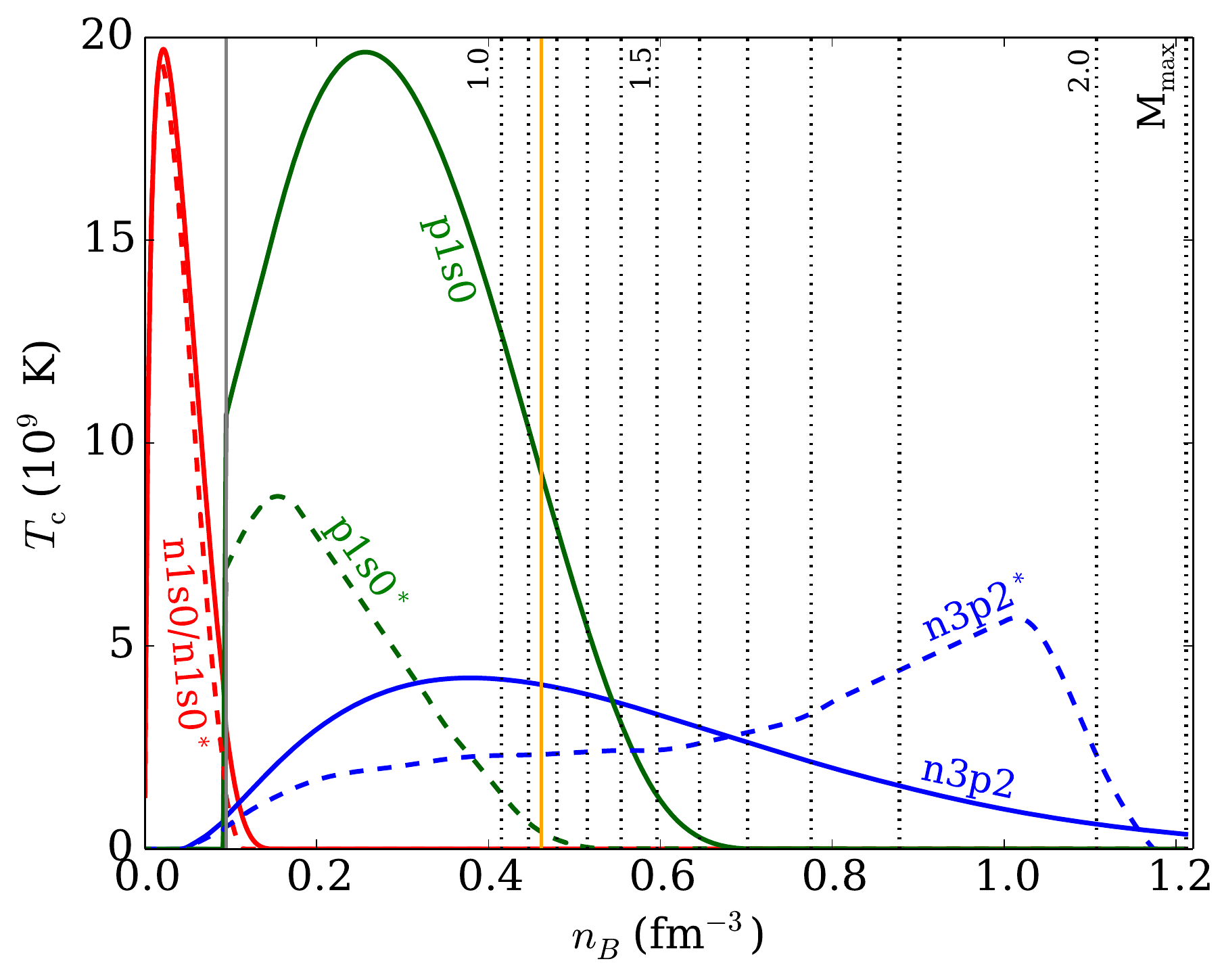}}
\vspace{-2mm}
\caption{(Color online)
Critical temperatures in NS matter for the BHF model
in the n1S0, p1S0 and n3P2 channels,
including ($^*$) or not effective mass effects.
The vertical dotted lines indicate the central density
of NS with different masses
$M/M_\odot=1.0,\ldots,2.0$, up to the maximum mass value.
DU onset occurs at the vertical solid (yellow) line
(see section~\ref{sect:micro}).}
\label{f:SPF}
\end{figure}

In Fig.~\ref{f:eos} we display the NS EOS obtained with the BHF model
for beta-stable and charge-neutral matter.
The upper panel shows pressure and energy density as functions of the baryon density,
whereas the middle panel shows proton and muon fractions.
The vertical red dotted line indicates the value of the density threshold
at which the DU process sets in, i.e.
$n_B=0.44\,\text{fm}^{-3}$, where $x_p=0.136$.
In the lower panel  we show the NS mass vs.~central density relation,
obtained by solving the standard Tolman-Oppenheimer-Volkov equations for the
NS structure.
We notice that the BHF model reaches a maximum mass slightly
above two solar masses $\mmax=2.01\,\ms$,
and thus is compatible with recent NS observations
\citep{Antoniadis2013,Fonseca},
and that the DU process can potentially operate in nearly all stars,
$M/M_\odot>1.10$.

This latter feature deserves a comment.
As was already mentioned,
our BHF calculations reproduce the empirical saturation parameters
of nuclear matter.
A relatively strong growth of the symmetry energy at supranuclear densities,
and the associated relatively low DU threshold density,
result from our microscopic calculations,
and we had no freedom to modify these quantities within our approach.

\subsection{Pairing gaps}

An essential ingredient for cooling simulations is the knowledge of the
1S0 and 3P2 pairing gaps for neutrons and protons in beta-stable matter,
which on one hand block neutrino processes and on the other hand open new ones
(see Section \ref{sect:micro}) \citep{YK01}.
In this paper we present calculations obtained using gaps computed consistently
with the EOS, i.e.,
based on the same NN interaction and using the same medium effects
(TBF and effective masses),
as shown in \cite{ourgaps}.
To be more precise,
and focusing on the more general case of pairing in the coupled 3P2 channel,
the pairing gaps were computed on the
BCS level by solving the (angle-averaged) gap equation
\citep{bcsp1,bcsp2,bcsp3,bcsp4,bcsp5,bcsp6}
for the two-component $L=1,3$ gap function,
\be
 \left(\!\!\!\begin{array}{l} \Delta_1 \\ \Delta_3 \end{array}\!\!\!\right)\!(k)
 = - {1\over\pi} \int_0^{\infty}\!\! dk' {k'}^2 {1\over E(k')}
 \left(\!\!\!\begin{array}{ll}
 V_{11} & V_{13} \\ V_{31} & V_{33}
 \end{array}\!\!\!\right)\!(k,k')
 \left(\!\!\!\begin{array}{l} \Delta_1 \\ \Delta_3 \end{array}\!\!\!\right)\!(k')
\label{e:gap}
\ee
with
\be
  E(k)^2 = [e(k)-\mu]^2 + \Delta_1(k)^2 + \Delta_3(k)^2 \:,
\ee
while fixing the (neutron or proton) density,
\be
 n = {k_F^3\over 3\pi^2}
 = 2 \sum_k {1\over 2} \left[ 1 - { e(k)-\mu \over E(k)} \right] \:.
\label{e:rho}
\ee
Here $e(k)$ are the BHF s.p.~energies, Eq.~(\ref{e:en}),
containing contributions due to two-body and three-body forces,
$\mu \approx e(k_F)$ is the chemical potential
determined self-consistently from Eqs.~(\ref{e:gap}--\ref{e:rho}),
and
\be
 V^{}_{LL'}(k,k') =
 \int_0^\infty \! dr\, r^2\, j_{L'}(k'r)\, V^{TS}_{LL'}(r)\, j_L(kr)
\label{e:v}
\ee
are the relevant potential matrix elements
($T=1$ and
$S=1$; $L,L'=1,3$ for the 3P2 channel,
$S=0$; $L,L'=0$ for the 1S0 channel)
with
\be
 V = V_{18} + \bar{V}_\text{UIX} \:,
\label{e:v23}
\ee
composed of two-body force and averaged TBF.
The relation between (angle-averaged) pairing gap at zero temperature
$\Delta \equiv \sqrt{\Delta_1^2(k_F)+\Delta_3^2(k_F)}$
obtained in this way and the critical temperature of superfluidity is then
$T_c \approx 0.567\Delta$.

Fig.~\ref{f:SPF} displays the various pairing gaps
as a function of baryonic density of beta-stable matter for the  BHF model.
In this way one can easily identify which range of gaps is active
in different stars,
whose central densities are shown by vertical dotted lines for given NS mass.
We stress that in \cite{ourgaps} no polarization corrections
\citep{pol1,pol,pol3,ppol3}
were taken into account,
which for the 1S0 channel are known to be repulsive,
but for the 3P2 are still essentially unknown;
and this might change the value of these gaps even by orders of magnitude
\citep{ppol1,ppol2,ppol3}.

In order to represent this uncertainty,
in the cooling simulations performed in \cite{Taranto2016}
we used the density dependence of the pairing gaps shown in Fig.~\ref{f:SPF},
but employing global scaling factors $s_p$ and $s_n$
for the p1S0 and n3P2 gaps, respectively.
One should notice that the n3P2 gaps shown in the figure are larger than those
currently employed in cooling simulations,
and that at the moment there exists no satisfactory theoretical
calculation of p-wave pairing that includes consistently all medium effects.

\subsection{Models of the crusts}

The structure of the NS crust depends on its formation scenario
\citep{HZ1990a,Zdunik2011}.
Consider first an isolated NS born in a core-collapse supernova explosion.
Then the crust is formed during the cooling of the very hot
and fully fluid proto-NS.
In the initial state,
the composition of the outer layer corresponds to nuclear equilibrium,
because $T>10^{10}\,$K and therefore nuclear reactions are sufficiently rapid.
The hot plasma crystallizes in the process of cooling.
A standard assumption is that during the process of cooling and
crystallization the plasma keeps the nuclear equilibrium.
Consequently, when matter becomes strongly degenerate,
the structure and EOS of the crust can be well approximated by
{\it cold catalyzed matter}
- the ground state (GS) of the matter at $T=0$.
The structure of the GS crust is calculated using
the energy density functional (EDF) based on the microscopic BHF theory of
nuclear matter,
applied previously for the NS core.
In this way we maintain a unified description of the whole NS interior
based on a microscopic theory of nuclear matter \citep{Sharma2015}.

The GS approximation is not valid for NS in qXRT.
Such NS formed their crust by accreting plasma from their companion star in
a low-mass X-ray binary.
The matter accreted onto the NS surface during $(10^8-10^9)\,$yr
compresses and pushes down the original pre-accretion crust.
During the active stages of a XRT the outermost layer of accreted
plasma undergoes thermonuclear flashes,
observed as X-ray bursts,
but the layers deeper than a few meters have $T<5\times 10^8\,$K
and therefore cannot keep nuclear equilibrium during compression.
On the contrary, with increasing depth and density,
the composition of the matter diverges from nuclear equilibrium,
because the temperature is too low for the thermonuclear fusion to proceed:
the fusion reactions are blocked by Coulomb barriers.
Consequently, the only nuclear reactions are electron
captures and neutron emissions and absorptions induced by compression.

Due to electron captures the neutron excess in nuclei increases with increasing
density,
and finally neutrons start to drip out from nuclei.
The neutron drip is triggered by the electron capture;
for our specific model this occurs at $6\times 10^{11}\,{\rm g\,cm^{-3}}$,
which is similar to the neutron drip density in the GS crust.
With increasing density,
electron captures trigger further neutron emission,
so that the mass number of nuclei decreases,
and they become more and more neutron rich.
Then, at densities greater than $10^{12}\,{\rm g\,cm^{-3}}$ the proton number
in the nuclei becomes so low ($Z\sim 10$) that pycnonuclear fusion of two
neighboring nuclei due to the (growing with density) zero-point motion of
nuclei around their lattice sites becomes possible on a timescale shorter
than the compression time.
This leads to the further neutronization of the crust matter.
Eventually, all the pre-accretion crust dissolves in the liquid core,
and a fully-accreted crust is formed.
Then, the layered structure of the crust ceases to evolve
and becomes quasistationary,
with matter elements moving inwards due to compression and undergoing exothermic
nuclear transformations.

The total heat release is calculated using a microscopic model
of the dense degenerate plasma,
and by following the nuclear evolution of an element of matter
consisting initially of X-ray ashes,
under quasistatic compression from
$10^7\,{\rm g\,cm^{-3}}$ to $10^{14}\,{\rm g\,cm^{-3}}$
(crust-core interface).
The microscopic model for a fully accreted crust is taken from \cite{HZ2008},
which calculated the EOS and distribution of deep crustal heating sources.
It is assumed that the X-ray bursts ashes consist of $A=56$ nuclei.
Note that the distribution of deep heating sources does not influence
the stationary thermal state of qXRT,
and only the total heat release is important.
The total deep crustal heating per one accreted nucleon is in our case
$Q_\text{DCH}=1.9\,$MeV (\citealt{HZ2003}).

\section{Thermal evolution of INS and qXRT}

\subsection{Thermal evolution of a INS}

Let us start by sketching the thermal evolution of a INS
(see \citealt{YP04,Page09} for a detailed review).
A proto-neutron star is formed in a supernova event with a high temperature
$T\sim 10^{11}\,$K.
The proto-neutron star becomes a NS when it gets transparent to the
neutrinos that are formed in its interior.
During $\sim 100$ years,
the crust stays hot due to its low thermal conductivity,
while the core cools by emission of neutrinos.
Therefore, the core and the crust cool independently
and the evolution of the surface temperature reflects the
thermal state of the crust and is sensitive to its physical properties
(see, e.g.,~\citealt{Fortin10} and references therein).
Then the core and the crust thermal evolutions couple.
The cooling wave from the core reaches the surface and the whole NS
cools by the emission of neutrinos, mainly from the core.
This is the neutrino cooling stage.
The evolution of the surface temperature depends mainly on the
physical properties of the core.
As the temperature in the interior of the NS continues to decrease,
the neutrino luminosity becomes comparable to the photon luminosity.
The NS then enters the photon cooling stage and the
evolution of the internal temperature is governed by the emission of
photons from the surface and is sensitive to the properties of the outer
parts of the star.

Currently, all INS for which the surface temperature has been measured
thanks to X-ray observations are at least $\sim 300$ years old
and they are all in the neutrino- and photon-cooling stages
(see, e.g.,~Table~1 in \citealt{BY15a}).
Therefore they have an isothermal interior, i.e.,
the redshifted internal temperature
$T_\text{i}(t)=T(t,r){\rm e}^\phi(r)$
is constant throughout the interior for densities larger than
$\rho_B\simeq 10^{10}\,$g cm$^{-3}$.
Here $T(r,t)$ is the local temperature and
${\rm e}^\phi(r)$ the metric function.
Thus the heat equation reads:
\be
 C(T_{\rm i})\frac{{\rm d} T_{\rm i}}{{\rm d} t} =
 -L_\nu^\infty(T_{\rm i})-L_{\gamma}^\infty(T_{\rm s}) \:.
\label{eqn:heatINS}
\ee
In this equation $T_{\rm s}(t)$ is the effective surface temperature
and the relation $T_{\rm s}(T_{\rm i})$ is determined by the model
of the heat blanketing envelope (see hereinafter).
$C(T_{\rm i})$ is the total specific heat integrated over the whole star
and the quantities $L_\nu^\infty$, $L_{\gamma}^\infty$
are respectively the redshifted total neutrino luminosity
and the surface photon luminosity detected by an observer at infinity:
\bea
 L_\nu^\infty(T_{\rm i}) &=&
 4\pi\int_0^R\!\!\! {\rm d}r {r^2{\rm e}^{2\phi}Q_\nu}/{\sqrt{ 1-2Gm/r}} \:,
\\
 \lgi(T_{\rm s}) &=&
 4\pi R^2\sigma_{\rm SB} T_{\rm s}^4 \big( 1-2GM/R \big) \:,
\eea
with $\sigma_{\rm SB}$ being the Stefan-Boltzmann constant,
$Q_\nu$ the neutrino emissivity,
$m(r)$ the gravitational mass enclosed in a sphere of radius $r$,
and $M=m(R)$ the gravitational mass of the star of radius $R$.
The temperature measured by an observer at infinity is then
$T_\mathrm{s}^\infty = {\rm e}^{\phi(R)} T_{\rm s}
= T_{\rm s} \sqrt{1-2GM/R}$.

\subsection{Microphysical ingredients}
\label{sect:micro}

The main contribution to the total specific heat $C(T_{\rm i})$ comes from
the NS core, i.e., from the neutrons, protons and electrons.
If the core is non-superfluid,
most of the specific heat comes from the neutrons.
If the neutrons and protons are superfluid, then below their superfluid critical
temperature $T_c$ their contributions become exponentially suppressed
and $C$ is dominated by the electrons.

Let us now review the various contributions to the neutrino emissivity
in the NS core,
which dominates the total neutrino luminosity $L_\nu^\infty$.
In a non-superfluid NS
the most powerful neutrino process is the so-called direct Urca (DU) process,
which is in fact the neutron $\beta$-decay followed by its inverse reaction:
\be
 n \rightarrow p + e^- + \bar{\nu}_e
 \qquad \textrm{and} \qquad
 p+e^- \rightarrow n+\nu_e \:.
\label{eqn:DU}
\ee
For this process, the neutrino emissivity varies as $T^6$.
However, the energy and momentum conservation imposes a density threshold
to this process \citep{LP91}.
For the EOS used in this work the DU process sets in at
$n_{\rm B}^{\rm DU}=0.44\,\text{fm}^{-3}$,
where the proton fraction $x_p=0.136$,
as indicated by the vertical red dotted line in Fig.~\ref{f:eos},
and thus the DU process operates in the central part
 of NS with masses larger than $\mdu=1.10\,\ms$,
corresponding to the yellow vertical solid line in Fig.~\ref{f:SPF}.

Various less efficient neutrino processes may be operating in the NS core
(see \citealt{YK01} for a review)
and dominate when the DU process is forbidden or strongly reduced.
The two main ones with an emissivity $Q_\nu\propto T^8$ are:
\begin{itemize}
\item the so-called modified Urca (MU) process:
\be
 n+N \rightarrow p + e^- + \bar{\nu}_e+N
 \qquad \textrm{and} \qquad
 p+e^-+N \rightarrow n+\nu_e+N \:,
\label{eqn:MU}
\ee
where $N$ is a spectator nucleon that ensures momentum conservation.
Nevertheless, since five degenerate fermions are involved instead of three,
the efficiency is significantly reduced as compared with the DU process.
\item the nucleon-nucleon bremsstrahlung:
\be
 N+N \rightarrow N+N+ \nu+\bar{\nu} \:,
\ee
with $N$ a nucleon and
$\nu$, $\bar{\nu}$ an (anti)neutrino of any flavor.
\end{itemize}

\begin{figure}
\vspace{-1mm}
\centerline{\includegraphics[scale=0.45]{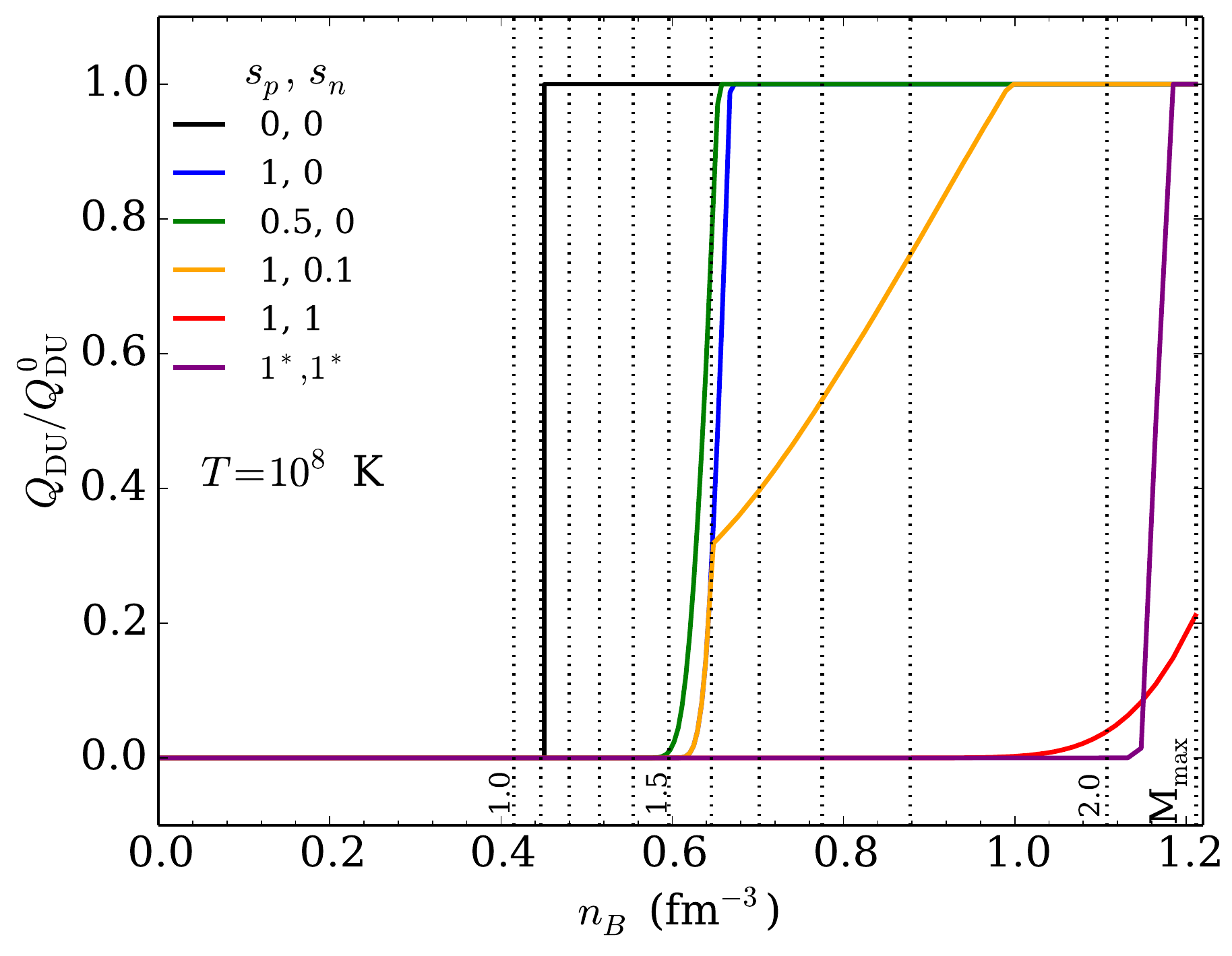}}
\vspace{-2mm}
\caption{(Color online)
DU ratio:
DU emissivity $Q_{\rm DU}$ for different superfluid models $(s_p,s_n)$
over DU emissivity $Q_{\rm DU}^0$ for non-superfluid matter,
at $T=10^8$ K.
The vertical dotted lines indicate the central density
of NS with different masses
$M/\ms=1.0,1.1,\ldots,2.0$,
up to $\mmax=2.01\ms$.}
\label{fig:DU}
\end{figure}

The effect of the neutron or proton superfluidity on the neutrino emissivity
is twofold.
On one hand,
when the temperature decreases below the critical superfluid
temperature of a given type of baryons, the neutrino emissivity of
processes involving a superfluid baryon is exponentially reduced.
For example, proton superfluidity in the core of a NS suppresses both Urca
processes (DU and MU) but does not affect the neutron-neutron
bremsstrahlung.

Fig.~\ref{fig:DU} shows,
as a function of the baryon number density of NS matter,
the ratio of the emissivity of the DU process for
various superfluid models over the DU emissivity in non-superfluid matter,
$Q_{\rm DU}^0$,
for a temperature $T=10^8\,$K.
For densities lower than the onset density $\nbdu=0.44\,\text{fm}^{-3}$,
or equivalently masses smaller than $\mdu=1.1\ms$
the ratio is set to zero and the DU process can not operate
at the center of the star.
In the case of non-superfluid matter (black curve),
the ratio for
$n_B>\nbdu$ (or at the center of stars with $M>\mdu$)
is by definition equal to 1 and the DU process is fully operative
in the central part of the NS.

Let us now turn on the proton superfluid without effective mass effect,
but not the neutron one;
this case corresponds to $s_p=1$ and $s_n=0$ (blue curve).
The proton critical temperature $T_c^p$ is smaller than $T=10^8\,$K
for densities larger than $\nbp(1,0)=0.67\,$fm$^{-3}$,
or masses above $\mp(1,0)=1.65\,\ms$.
Hence the DU emissivity is not affected by the superfluidity
for such densities and in the center of NS with such masses,
and the DU ratio is equal to 1.
For $\nbdu<n_B<\nbp(1,0)$
or equivalently $\mdu<M<\mp(1,0)$ however,
the DU process is allowed and the protons are superfluid at $T=10^8\,$K.
These being involved in the reaction in Eq.~(\ref{eqn:DU}),
the DU emissivity and thus the ratio is exponentially reduced.
Consequently the DU ratio is no longer a step-like function,
but a smooth function of the density or the NS mass:
the DU process is turned off below $\mdu$,
is equal to the one of non-superfluid matter
at the center of NS with masses bigger than $\mp(1,0)$,
above which the protons are no longer superfluid,
and is reduced due to superfluid effects in between the two boundaries.

The superfluid effects are similar if we now artificially divide
the proton critical temperature by a factor of 2,
while keeping the neutrons non-superfluid,
($s_p=0.5$, $s_n=0$, green curve).
Only the density and mass at which $T_c^p<10^8\,$K differ:
$\nbp(0.5,0)=0.66\,$fm$^{-3}$, $\mp(0.5,0)=1.63\,\ms$.

Let us now consider superfluid protons (with no scaling) and scale the neutron
superfluidity by a factor 0.1
($s_p=1$, $s_n=0.1$, yellow curve).
Then the density for which the proton critical temperature
is smaller than $10^8\,$K is $\nbp(1,0.1)=\nbp(1,0)$,
hence $\mp(1,0.1)=\mp(1,0)$,
and for the neutron $\nbn(1,0.1)=1.0\,$fm$^{-3}$,
which is larger than $\nbp(1,0.1)$ ($\mn(1,0.1)=1.97\,\ms$).
If two species involved in the DU process are superfluid,
then the exponential reduction is roughly speaking dominated
by the stronger superfluidity \citep{YK01},
in this case the proton one.
Consequently for $\nbdu<n_B<\nbp(1,0.1)$  
the ratio is similar as for the case $s_p=1$ and $s_n=0$.
Then for $\nbp(1,0.1)<n_B<\nbn(1,0.1)$   
the DU process is not suppressed because of the proton superfluidity,
but because of the neutron one.
Hence the ratio is still smaller than 1.
Finally when $n_B>\nbn(1,0.1)$,          
the DU process is not affected anymore and the ratio equal to 1.

If we now consider the two unscaled models for the proton and neutron
superfluidity without medium effects (i.e., $m_i^*=m_i$),
$s_p=1$ and $s_n=1$ (red curve),
the neutron critical temperature remains larger than $10^8\,$K
up to densities comparable to the central density of the NS with the
maximum mass $n_B^c(\mmax)$.
Thus the DU process is strongly reduced
because of the proton and neutron superfluidity and the
DU ratio is much smaller than 1 for any NS mass.
Similar observations can be made for the case
of the proton and neutron superfluidities
including the effective mass effects ($m_i$ replaced by $m_i^*$),
$s_p=1^*$ and $s_n=1^*$ (darkred curve),
except that neither the protons nor the neutrons are superfluid above
$\nbp(1^*,1^*)=1.19\,$fm$^{-3}$, $\mp(1^*,1^*)=2.005\,\ms$,
and thus the ratio is equal to 1.

On the other hand,
the pairing of baryons initiates a new type of neutrino processes
called the pair breaking and formation (PBF) processes.
The energy is released in the form of a neutrino-antineutrino pair
when a Cooper pair of baryons is formed.
The process starts when $T<T_c$,
is maximally efficient when $T\sim 0.8\,T_c$,
and is exponentially suppressed for $T\ll T_c$ (\citealt{YK01}).
For both 1S0 and 3P2 pairings,
the emission of neutrinos by the PBF process can occur
through two different channels:
the axial and vector channels \citep{Page09}.
Since the vector part of the PBF process is strongly suppressed
\citep{LP06a,LP06b},
the axial part is in fact the main contributor to the PBF neutrino emissivity.
In addition, as pointed out recently in \cite{L16} and references therein:
(a) for the 1S0 pairing of baryons,
in our case for the protons in the core,
the neutrino losses in the axial channel are very small
and can be neglected in the calculations;
(b) for the n3P2 pairing in the core,
taking into account the anomalous contributions in the axial channel
leads to an emissivity four times smaller than the one used in previous works,
e.g., in \citet{P09,Taranto2016}.

Yet, as the n3P2 gap extends up to very large density,
see Fig.~\ref{f:SPF},
superfluidity provides an efficient means to block the DU cooling process,
in particular the n3P2$^*$ model taking into account medium effects.
The price to pay is an enhanced PBF cooling rate
close to the critical temperature in that case.
Also the p1S0 gap might persist throughout NS
up to about 1.6 $\ms$ (for pure BCS gaps),
and at this threshold the DU process will invariably set in and cool the
heavier stars very rapidly,
if neutrons are not superfluid.

\subsection{Heat-blanketing envelope}

In a few hundred years the redshifted temperature inside a newly born NS
for density larger than
$\rho_{\rm b} \simeq 10^{10}\,$g cm$^{-3}$
becomes uniform due to the high thermal conductivity.
However, in the NS atmosphere the heat transport is dominated by the photons
and in between there exists a thin layer,
which has a low thermal conductivity,
since the electrons are not highly degenerate
and the high density strongly prevents photon transport.
This results in high temperature gradients in the envelope,
that is few hundred meters thick.

Therefore, a variety of models
are devoted solely to the precise modeling of the heat-blanketing envelope,
in the plane-parallel and stationary approximation,
resulting in a relation between the surface temperature $T_{\rm s}$
and the temperature $T_{\rm i}$ at $\rho_{\rm b}$.
Some models consider various compositions for the envelope and in particular,
different abundances of light elements such as hydrogen, helium, carbon
resulting from the accretion of matter \citep{PC97}
or binary mixtures of these \citep{BF16}.
The higher the abundances of light elements,
the higher the surface temperature for a given $T_{\rm i}$.
As these abundances are unknown for a given NS,
we will consider in the following two limiting cases corresponding
to the absence of light elements (non-accreted envelope)
and a maximum amount of them (fully accreted envelope),
and use the non-magnetic models from \cite{PY03}.

\subsection{Thermal evolution of qXRT}

qXRT contain NS in binary systems,
which accrete matter from a companion star during short active phases
with a high luminosity followed by long period of quiescence
with zero or strongly reduced accretion and low luminosity.
Here, we will model the quiescent phase.
The accreted matter sinks gradually in the interior of the NS
and undergoes a series of nuclear reactions
(beta captures, neutron absorption and emission and pycnonuclear fusions).
These reactions release some heat,
$Q_{\rm DCH}\sim 2\,$MeV per accreted nucleon,
which propagates into the whole NS,
inwards heating the core and outwards emitted in the form of photons
at the surface.
This is the so-called deep crustal heating.
A state of thermal equilibrium with a constant $T_{\rm i}$ throughout the star
is reached in $\sim 10^5$ years as the episodes of heating due to the accretion
followed by quiescence and cooling through the emission of neutrinos proceed
(\citealt{YL03,YL04}).
The internal temperature $T_{\rm i}$ of such a star is then determined
by the following equation:
\be
 L_{\rm DCH}^\infty(\dot{M}) =
 L_\nu^\infty(T_{\rm i}) + L_{\gamma}^\infty(T_{\rm s})
\label{eqn:heatANS}
\ee
with $L_\nu^\infty$ and $L_{\gamma}^\infty$
being the neutrino and photon luminosities described before.
The deep crustal heating power $L_{\rm DCH}^\infty$ is given by:
\be
 L_{\rm DCH}^\infty(\dot{M}) =
 {\rm e}^{\phi}\frac{Q_{\rm DCH}}{m_{\rm N}}\dot{M}
\ee
with $m_{\rm N}$ the atomic mass unit and $\dot{M}$ the mean accretion rate,
assumed to be constant,
averaged over periods of accretion and quiescence.

\subsection{Cooling and heating curves}

In the following we make use of the one-dimensional code called
\texttt{NSCool}\footnote{freely available online:
{\url{http://www.astroscu.unam.mx/neutrones/NSCool/}}}
based on an implicit scheme developed by \cite{Henyey64},
suitable for the study of spherically symmetric problems,
that employs the Newton-Raphson method to solve the heat equation.
We then obtain for a NS of given mass and composition of the envelope
so-called {\em cooling curves} of INS showing equivalently
$L_{\gamma}^\infty$ as a function of the NS age $t$,
and {\em heating curves} of qXRT relating the $L_\gamma^{\infty}$
in quiescence to the estimated time-averaged accretion rate
$\dot{M}$ \citep{YL03}.
We will compute such cooling and heating curves for various superfluid models
using the gaps calculated with the same nuclear interaction model as the EOS.

Note that we do not include the effect of the magnetic field $B$
on the thermal evolution and on the thermal transport
in the envelope \citep{PY03}.
On one hand the magnetic field is likely to affect the cooling of INS
(see, e.g.,~\citealt{AP08,PM09}),
as it is estimated that they have $B\gtrsim 10^{12}\,$G;
however this is beyond the scope of the present paper.
On the other hand,
for qXRT its influence on cooling and heating can be neglected
as these objects have $B\simeq (10^8-10^9)\,$G
as a result of the accretion-induced decay of the magnetic field.

\begin{figure*}
\includegraphics[width=\columnwidth]{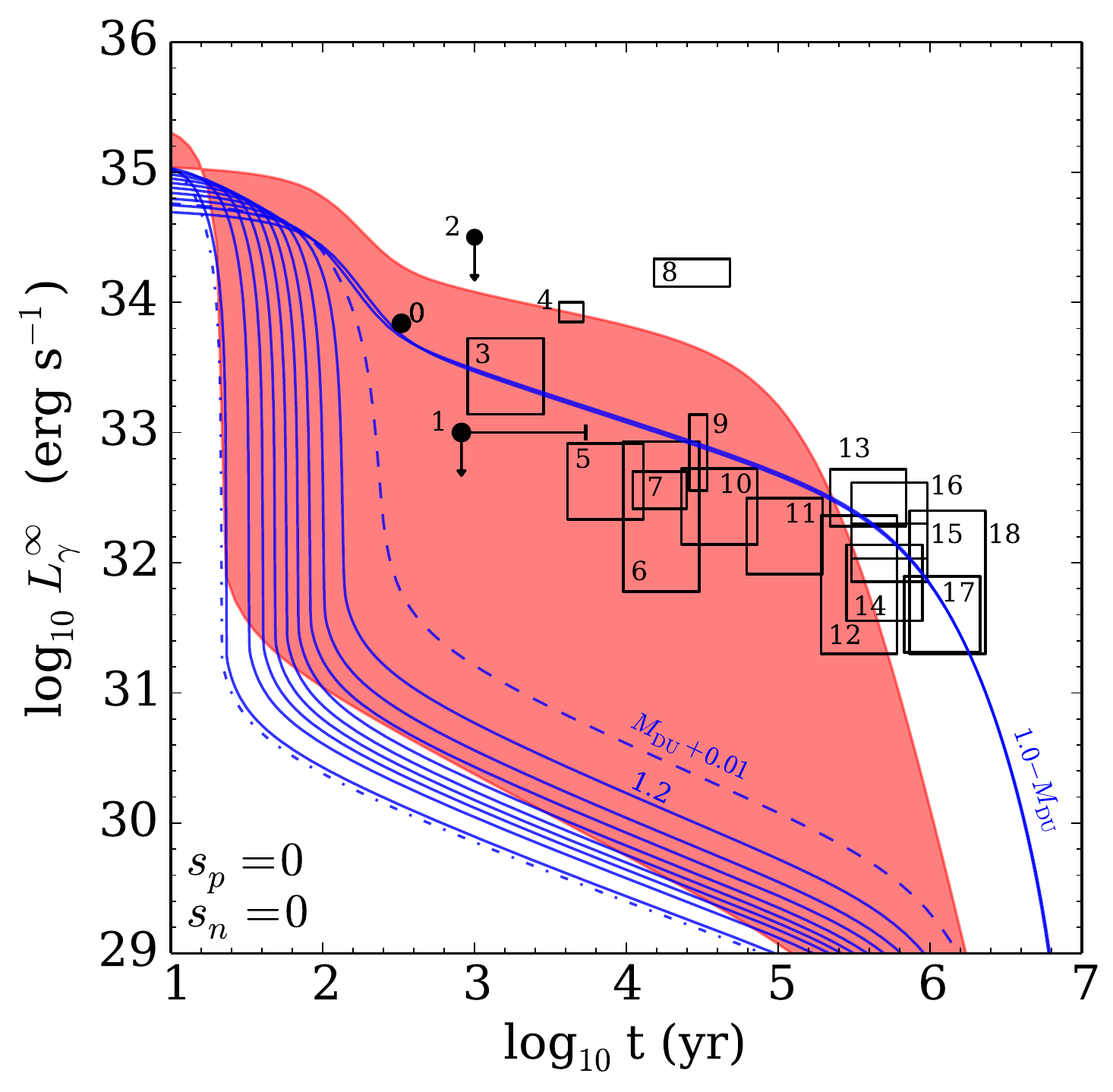}
\includegraphics[width=\columnwidth]{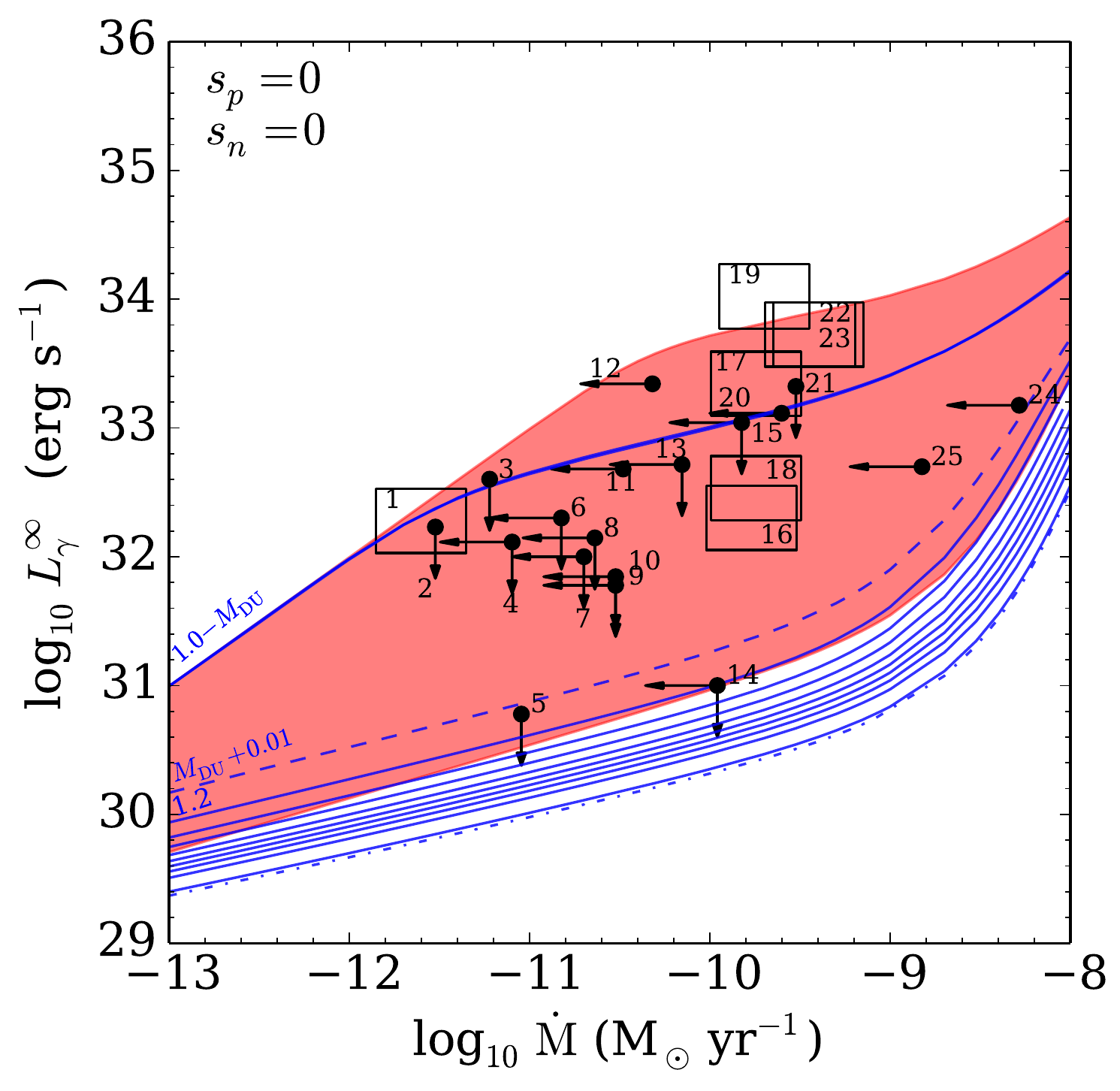}
\caption{
Thermal states of INS (left) and qXRT (right) for non-superfluid matter
($s_p=0$, $s_n=0$).
For a non-accreted (Fe) envelope, thermal states are plotted in blue
for $M/\ms=1.0, 1.1, 1.2, \ldots, 2$ (solid lines), for
$M=\mmax$ (dot-dashed line) and $M=\mdu+0.01\,\ms=1.11\,\ms$ (dashed).
For a fully-accreted envelope,
the red contour corresponds to the area covered by the thermal states
for $1.0\,\ms<M<\mmax$.
\newline
Observational data for INS:
0 - CasA NS,
1 - PSR J0205+6449 (in 3C58),
2 - PSR B0531+21 (Crab),
3 - PSR J1119$-$6127,
4 - RX J0822$-$4300 (in PupA),
5 - PSR J1357$-$6429,
6 - PSR B1706$-$44,
7 - PSR B0833$-$45 (Vela),
8 - XMMU J1731$-$347,
9 -  PSR J0538+2817,
10 - PSR B2334+61,
11 - PSR B0656+14,
12 - PSR B0633+1748 (Geminga),
13 - PSR J1741$-$2054,
14 - RX J1856.4$-$3754,
15 - PSR J0357+3205 (Morla),
16 - PSR B1055$-$52,
17 - PSR J2043+2740,
18 - RX J0720.4$-$3125.
\newline
For qXRT:
1 - IGR 00291+5934,
2 - XTE J1814$-$338,
3 - XTE J1751$-$305,
4 - XTE J1807$-$294,
5 - SAX J1808$-$3658,
6 - SAX J18104$-$2609,
7 - XTE J0929$-$314,
8 - XTE 2123$-$058,
9 - NGC6440 X2,
10 - EXO 17474$-$214,
11 - Cen X$-$4,
12 - 4U1730$-$22,
13 - 2S1803$-$245,
14 - 1H 1905+000,
15 - Terzan 1,
16 - MXB 1659$-$29,
17 - RX J1709$-$2639,
18 - NGC 6440 X1,
19 - SAX J1750.8$-$2900,
20 - 1M 1716$-$315,
21 - Terzan 5,
22 - 4U 1608$-$522,
23 - Aql X$-$1,
24 - 4U 2129+47,
25 - KS 1731$-$260.}
\label{fig:000}
\end{figure*}

\begin{figure}
\includegraphics[width=\columnwidth]{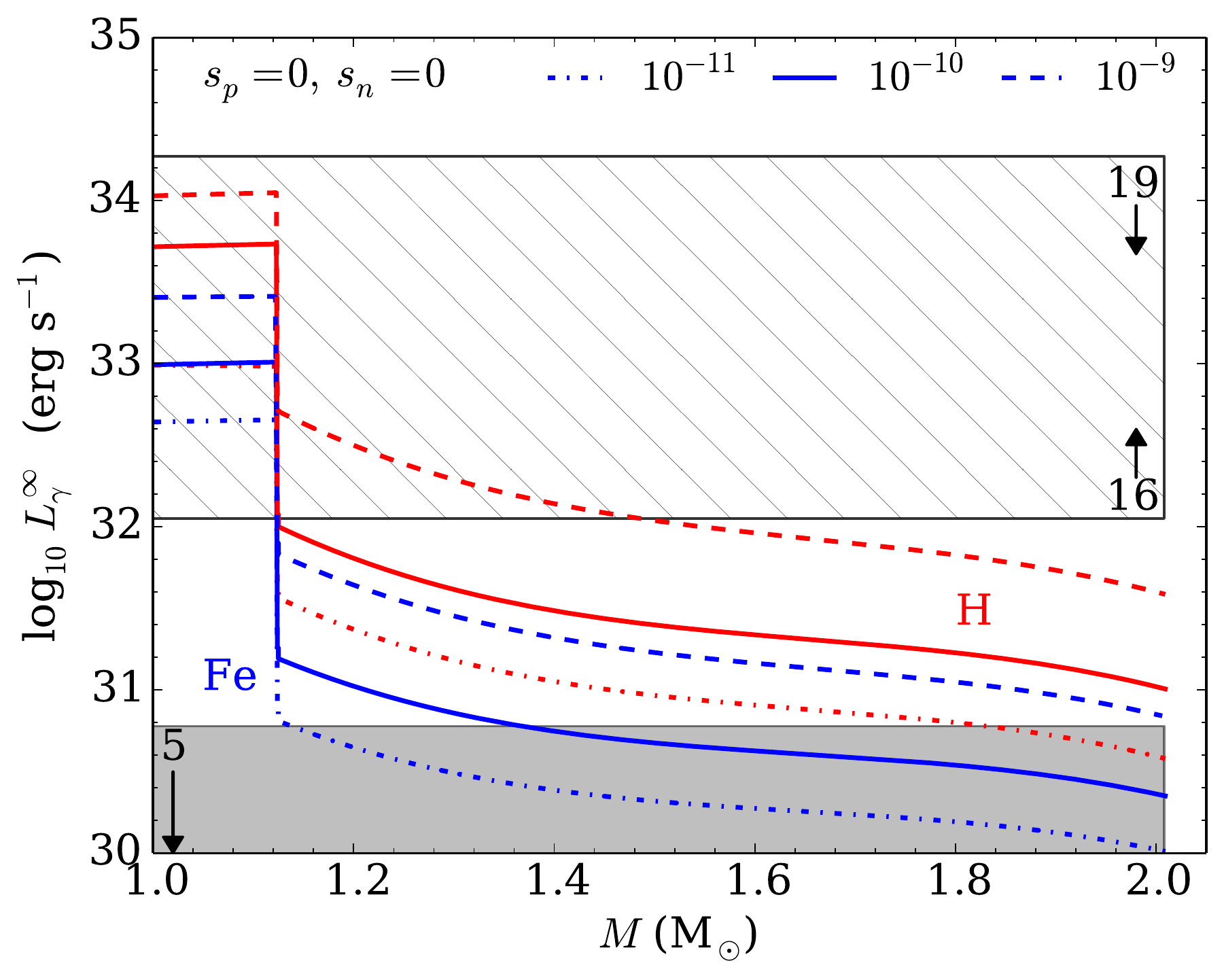}
\caption{
Surface photon luminosity $\lgi$ versus mass $M$
of a selection of qXRT for
$\dot{M}/\ms=10^{-(11,10,9)}\,$yr$^{-1}$.
Non-superfluid matter is considered ($s_p=0$, $s_n=0$),
together with two envelope models:
non-accreted (Fe) in blue and fully-accreted (H) in red.
The hatched region indicates the range of luminosities covered by
the objects 16, 17, 18, 19, 20, 22, 23 with relatively well determined
accretion rate and luminosity.
The grey strip corresponds to the low-luminosity object 5.}
\label{fig:000m}
\end{figure}

\begin{figure*}
\includegraphics[width=\columnwidth]{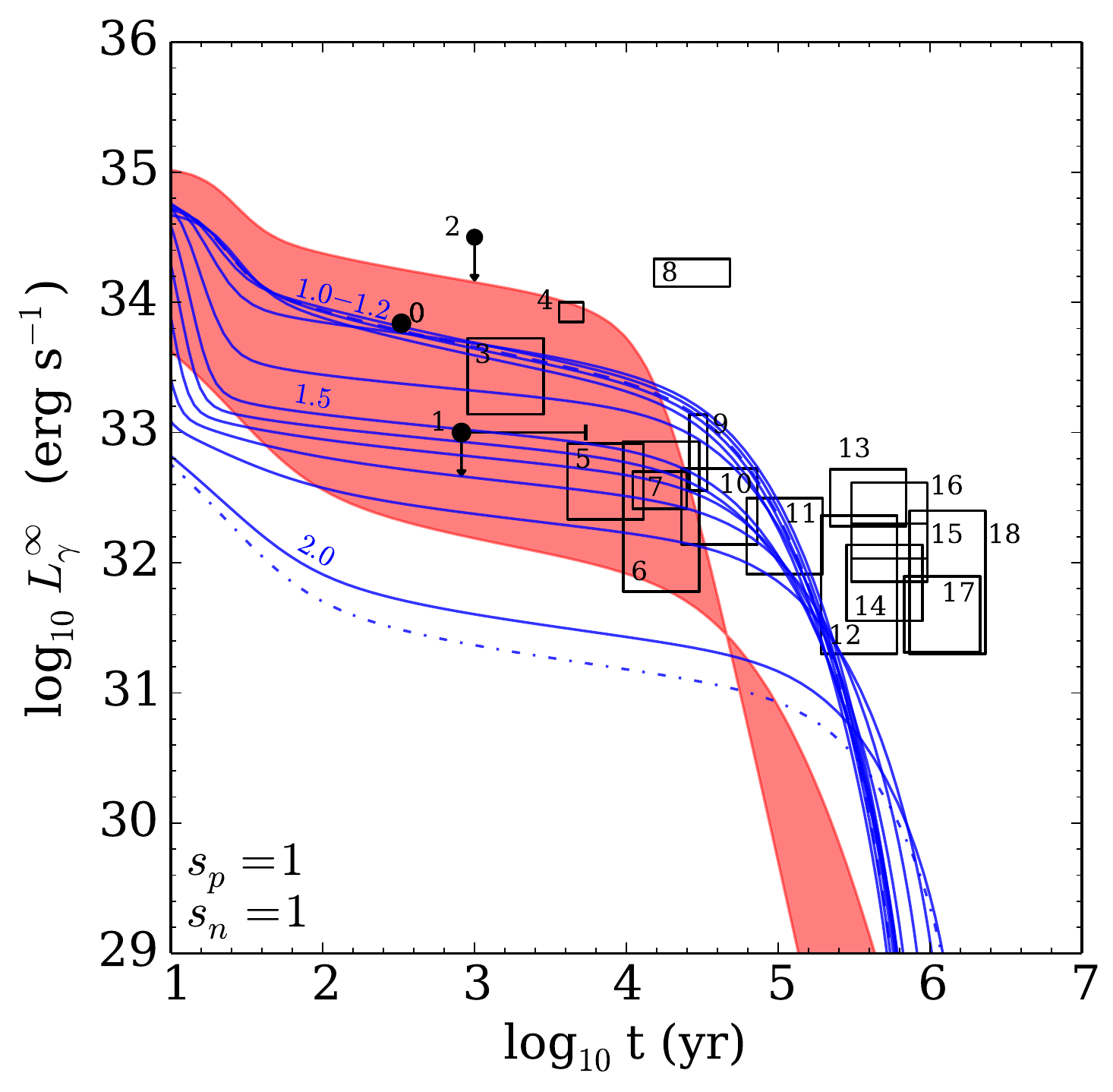}
\includegraphics[width=\columnwidth]{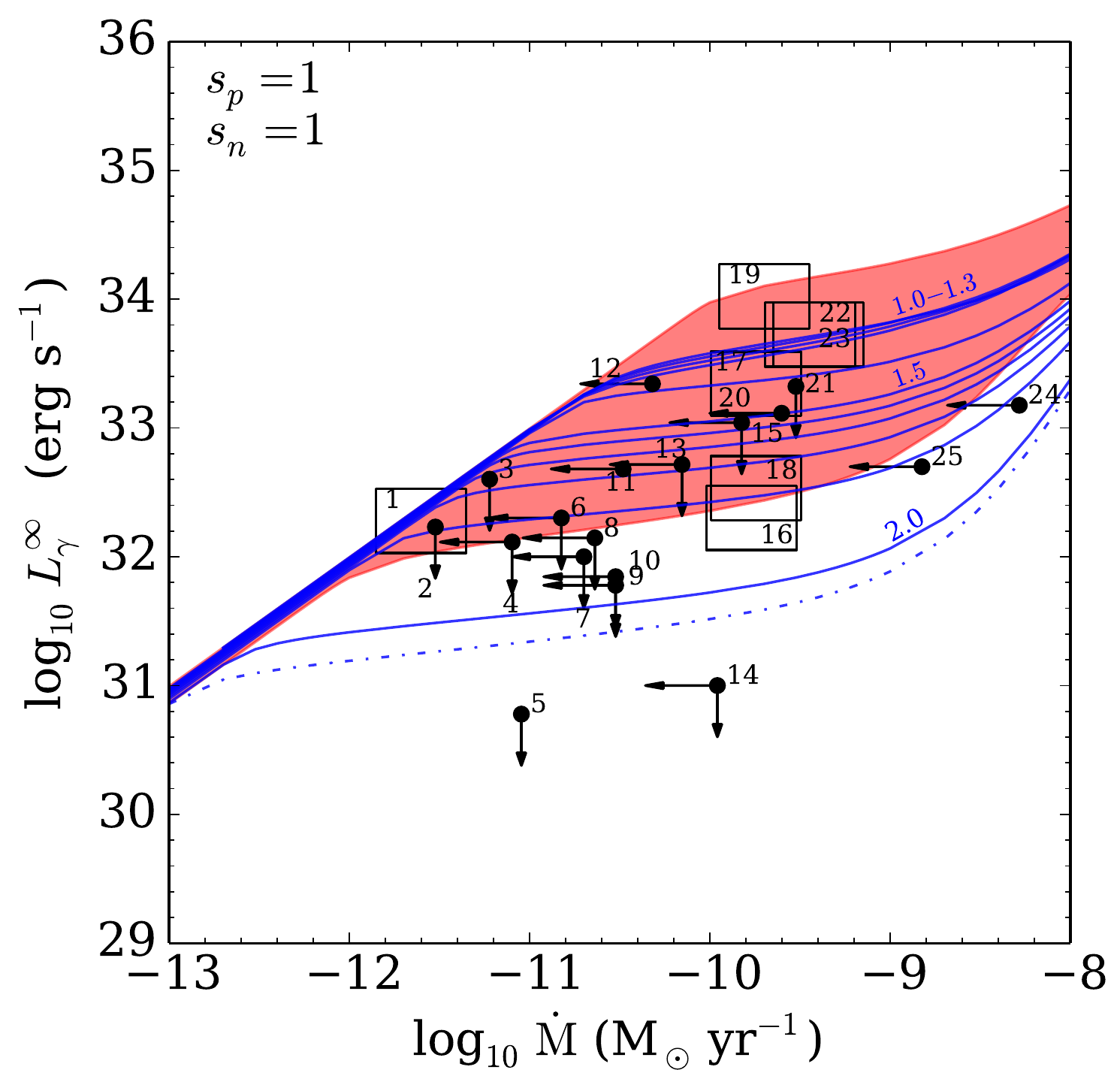}
\caption{
Thermal states of INS and NS in qXRT for superfluid NS matter
with gaps calculated without including the effective mass effects:
$s_p=1$, $s_n=1$.}
\label{fig:111}
\end{figure*}

\begin{figure*}
\includegraphics[width=\columnwidth]{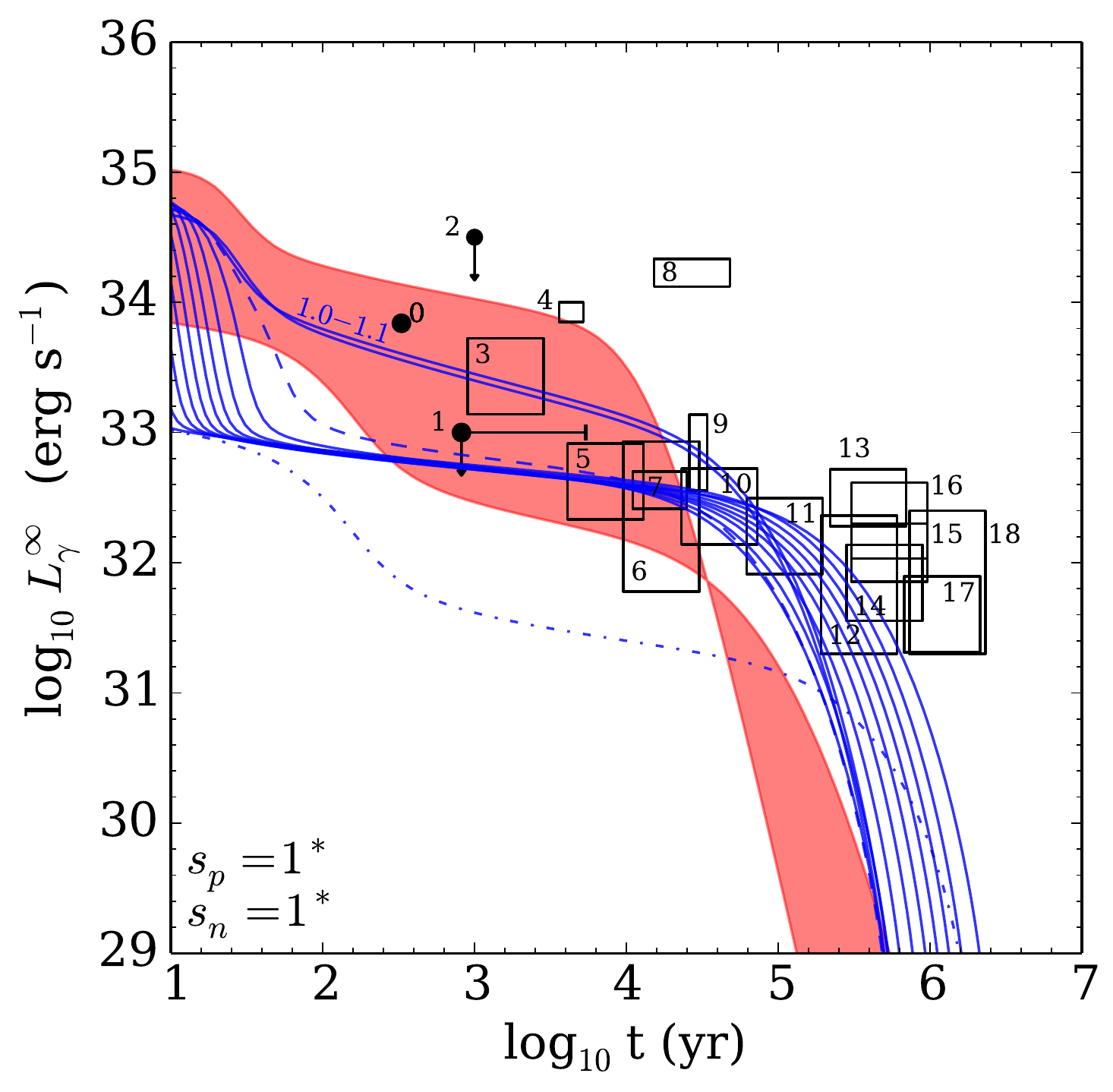}
\includegraphics[width=\columnwidth]{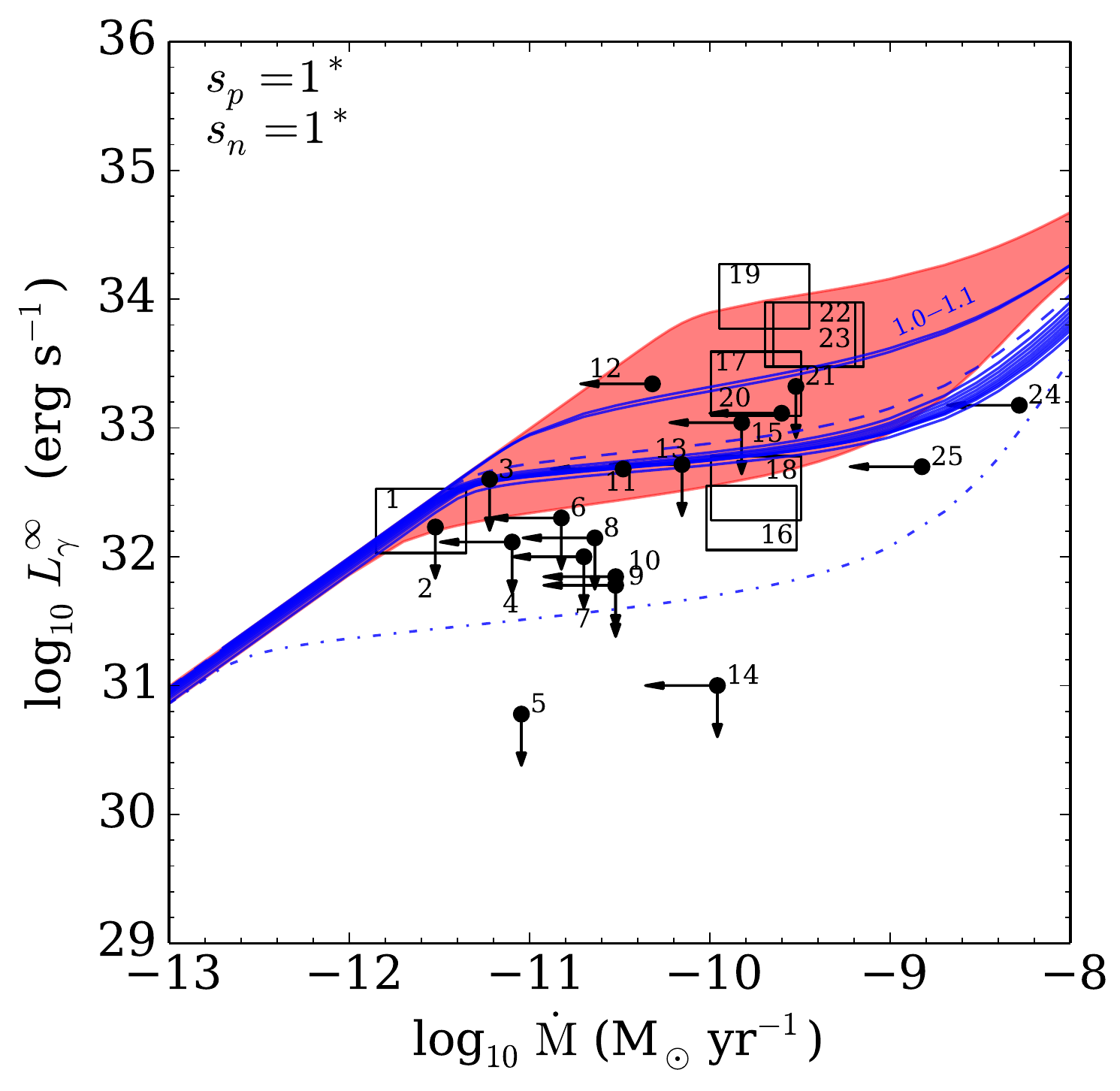}
\caption{
Thermal states of INS and NS in qXRT for superfluid NS matter
with gaps calculated including the effective mass effects:
$s_p=1^*$, $s_n=1^*$.}
\label{fig:1s1s1s}
\end{figure*}

\begin{figure*}
\includegraphics[width=\columnwidth]{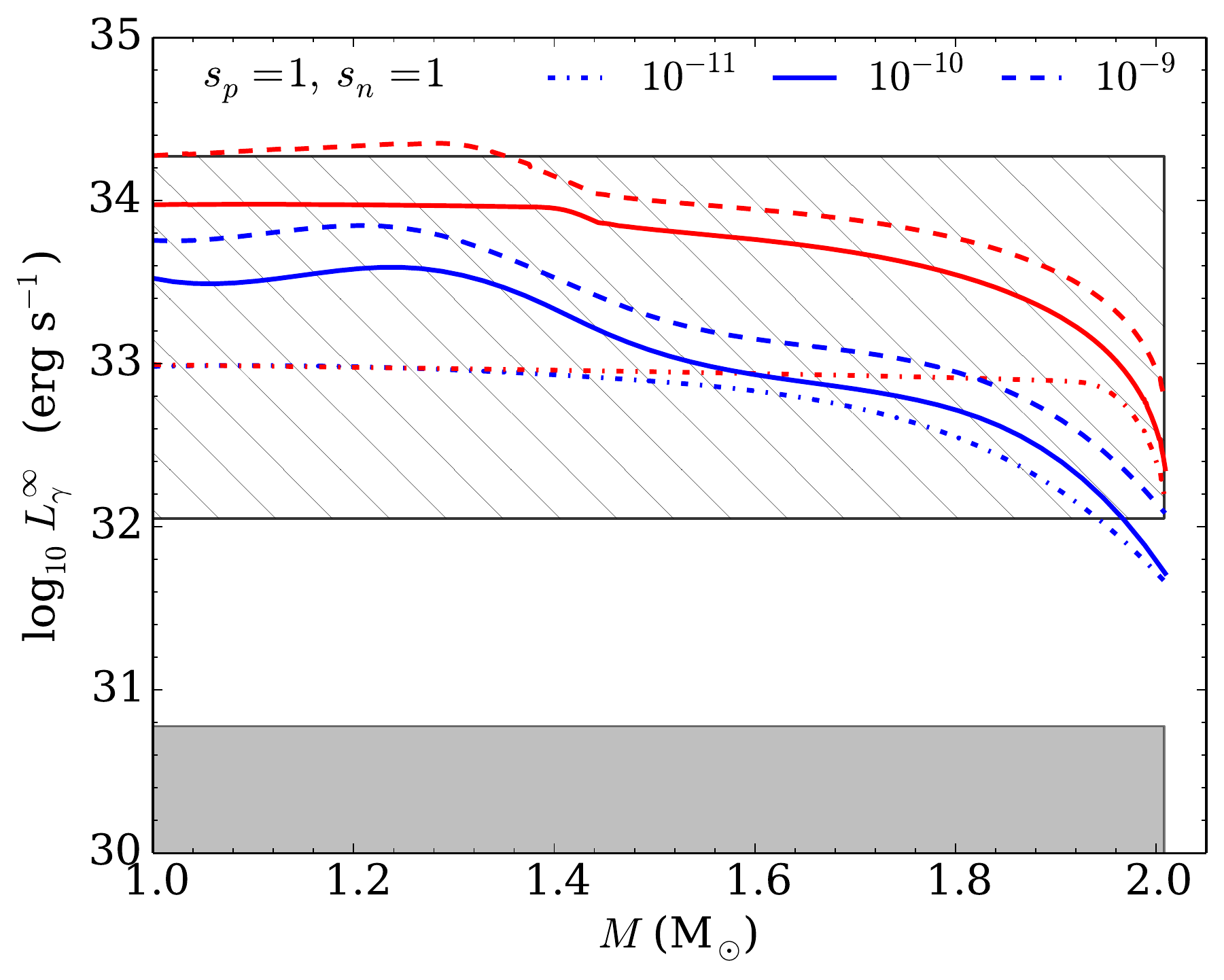}
\includegraphics[width=\columnwidth]{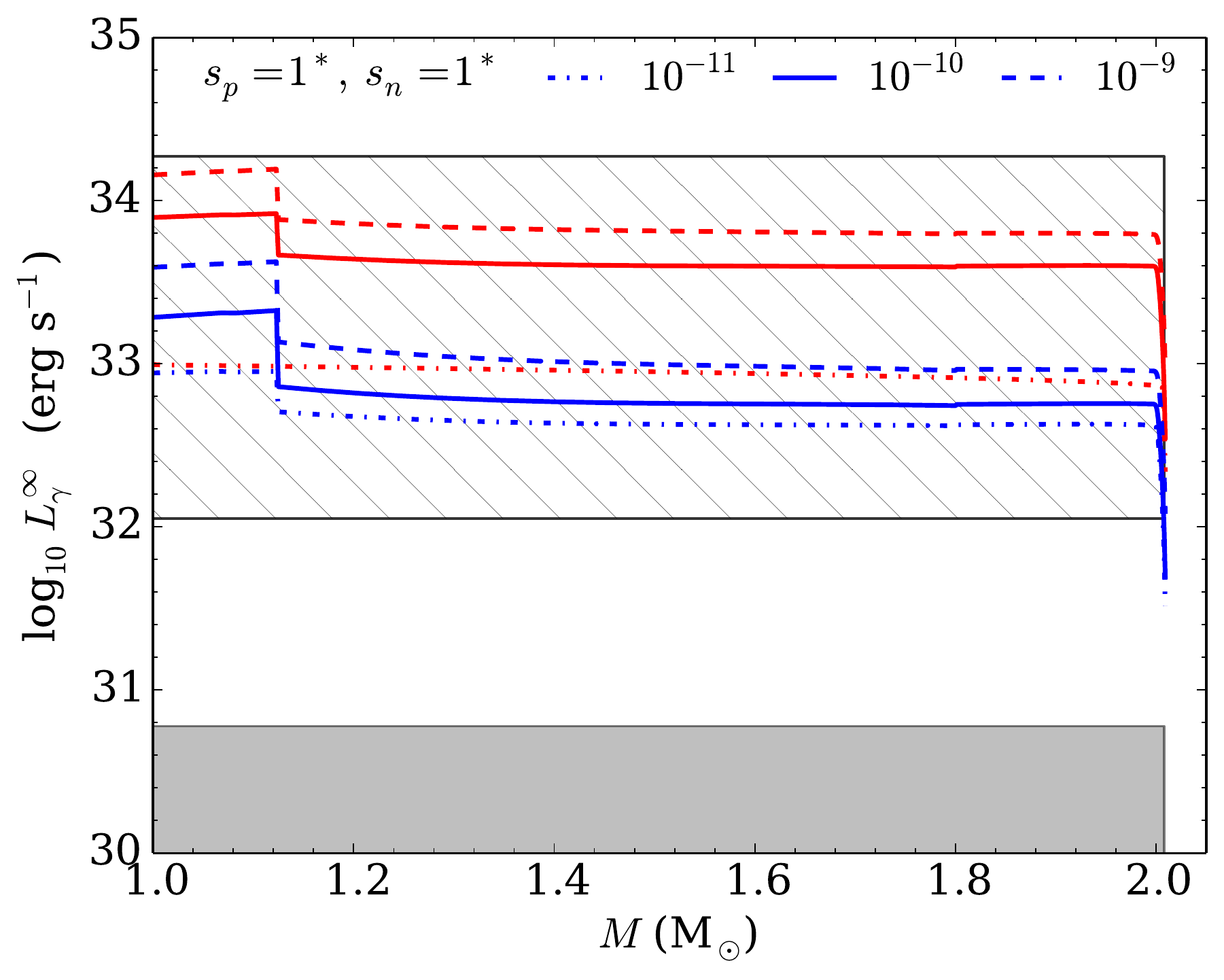}
\caption{
Surface photon luminosity $\lgi$ versus mass $M$
for a selection of qXRT.
Model with superfluid gaps calculated without (with) effective mass effects
on the left (right).}
\label{fig:111m}
\end{figure*}

\section{Results}

\subsection{Non-superfluid matter}

Let us first consider the thermal states of INS and qXRT for the case of
non-superfluid matter, i.e.,
for $s_p=0$, $s_n=0$, presented in Fig.~\ref{fig:000}.
Cooling and heating curves are calculated for masses
ranging from $1.0\,\ms$ to $M=\mmax$ for two limiting models of atmosphere
(non-accreted in blue and fully accreted in red).
Observational data for INS and qXRT from \cite{BY15a} and references therein
are also plotted.

However, as noted in the same reference,
it should be pointed out that in many cases the distance to the object,
the composition of its atmosphere, thus its luminosity,
and its age or average accretion rate are rather estimated than measured.
Thus in these cases, we use large error bars (a factor 0.5 and 2)
to reflect this uncertainty
and do not intend to constrain the mass of any object.
We point out that we do not model the thermal evolution of the CasA NS,
as it was already performed in \cite{Taranto2016}.
Note that the direct observation of its cooling is questioned,
see \cite{PP13,EH13}, but also \cite{Ho15}.
Two objects are particularly constraining:
1) XMMU J1731$-$347 \citep{XMMUb},
a middle-aged and yet very hot INS,
number 8 in Fig.~\ref{fig:000} left,
shown to have a carbon atmosphere
and whose mass has been constrained to be larger than $1.4\,\ms$
\citep{XMMUa},
but see the discussion about the limitations of this result in the reference;
2) SAX J1808$-$3658 \citep{SAX},
a low-luminosity qXRT with a well-constrained accretion rate,
number 5 in Fig.~\ref{fig:000} right,
which has been shown to have a relatively low mass $M\lesssim 1.7\,\ms$
by modeling its pulse shapes observed in X-ray \citep{SAXb}.

In a NS with $M<\mdu$,
the DU process is turned off and consequently the total neutrino emissivity
is orders of magnitude smaller than for a NS with a mass above the DU threshold.
Consequently the former has at a given age or accretion rate a higher luminosity
than the latter.
Note that on one hand all NS with $M<\mdu$ have a small neutrino emissivity,
hence their cooling and heating curves are nearly indistinguishable
on the scale of Fig.~\ref{fig:000}.
On the other hand,
for objects with $M>\mdu$,
the larger is the mass and thus the bigger is the central region of the star
where the DU process operates,
the lower is the luminosity
and the cooling/heating curves are no longer superimposed.
This behavior can clearly been seen in Fig~\ref{fig:000},
where cooling and heating curves are plotted for various masses.

Comparing the influence of the envelope model on INS,
for a given age in the neutrino-cooling stage,
a model obtained for a fully-accreted one has a higher luminosity
than one with a non-accreted envelope.
This originates from the fact that envelopes with a higher amount
of light elements have higher surface temperatures
and thus surface photon luminosities.
Hence models with a fully-accreted envelope will enter earlier
the photon cooling stage and then cool faster than models
with a non-accreted envelope.
In the case of qXRT,
at a given accretion rate
models with a fully-accreted envelope have larger luminosities
than the ones with a non-accreted envelope.

Thermal states of INS and qXRT obtained for non-superfluid matter
with two limiting envelope models can explain almost all the observational data.
Some objects like the INS~4 or the qXRT 12, 17, 19, 22, 23
require a fully-accreted envelope model and others, INS 17, 18,
a non-accreted one.
All the others are compatible with both envelope models.
The INS~2 and 8 appear challenging for the model since they are hot (luminous)
and young or middle-aged.
Among the qXRT, the objects 16, 17, 18, 19, 20, 22, 23
are particularly interesting as their luminosity and accretion rate
are rather well constrained,
as well as the object~5,
which has a well-constrained accretion rate and a very low luminosity.

In the case of non-superfluid matter,
the DU threshold is a step-like function.
Hence for $M<\mdu$ the DU process is turned off and for masses just above
the threshold is turned on and fully operating.
As a consequence, all 18 INS but the 2 too hot objects 2 and 8
would have masses between $1.0\,\ms$ and $\mdu=1.1\,\ms$.

In Fig.~\ref{fig:000m},
the relation between the NS mass and the luminosity of qXRT
is plotted for three different accretion rates:
$\dot{M}/\ms=10^{-(11,10,9)}\,$yr$^{-1}$
and the two envelope models.
In addition the hatched region delimits the range of luminosities
covered by the objects 16, 17, 18, 19, 20, 22, 23
with rather well-constrained observational data
and the grey area indicates the luminosity range of the least luminous object 5.
The range of accretion rates has been chosen to cover the ones
of all of these 8 objects.
The figure indicates that while the qXRT 16, 17, 18, 19, 20, 22, 23
all have masses $M\simeq(1.0-1.1)\,\ms$,
the object~5 requires larger neutrino losses
and thus the DU process to be fully operating
and the NS mass to be large enough in order to make the luminosity low enough.
Thus while the DU process does not appear required to explain
the thermal states of INS,
the low-luminosity qXRT SAX J1808$-$3658 can only be explained
if this efficient neutrino process is turned on already at not too low masses
(to ensure a large enough region where the process is operating and
thus producing sufficiently strong neutrino losses).

A striking feature of the relation $\lgi(M)$ in Fig.~\ref{fig:000m}
is its step-like behavior,
which is a direct consequence of the DU process having the same property.
It implies that for a very wide range of luminosities
$\lgi\sim(10^{31}-10^{33})\,$erg\,s$^{-1}$,
which is observationally relevant,
NS have a unique mass $M\simeq \mdu$.
In other words,
the thermal states of low-mass NS (with $M<\mdu$) weakly depend on the NS mass.
This feature appears difficult to reconcile with the current mass measurements
\citep{masses,OF16} with masses ranging from $\sim 1.0$ to $2.0\,\ms$,
as also pointed out in \cite{BY15a} using a statistical approach
to the INS and qXRT thermal states.
A solution to this problem studied in \cite{BY15a,BY15b}
consists in broadening of the DU threshold,
making it no longer a step-like function.
In the following we study a possible origin of this broadening
due to the inclusion of the proton and neutron superfluidity,
based on our model of consistent EOS and superfluid gaps.

\subsection{Consistent gaps and EOS}

\begin{figure*}
\includegraphics[width=\columnwidth]{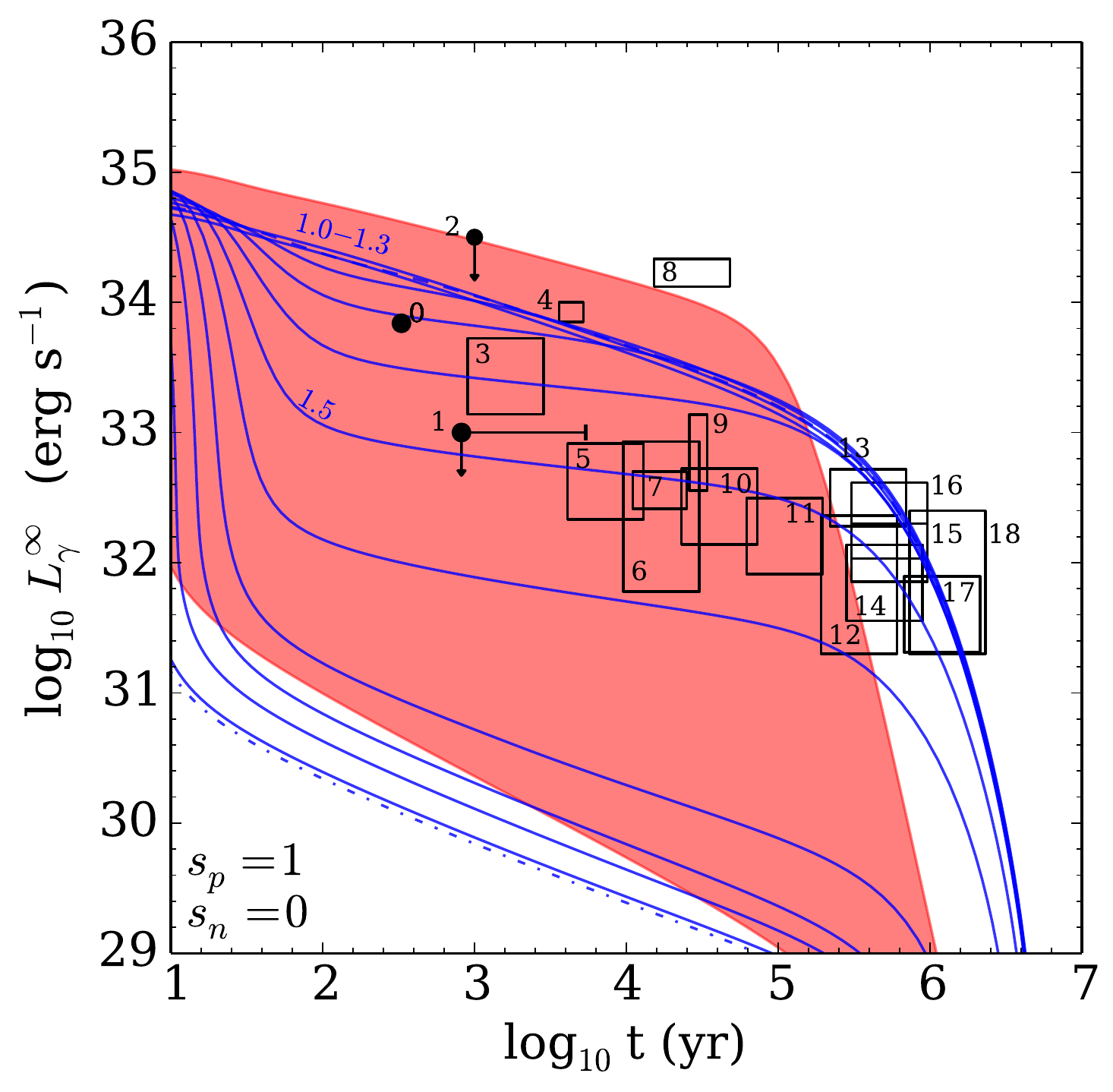}
\includegraphics[width=\columnwidth]{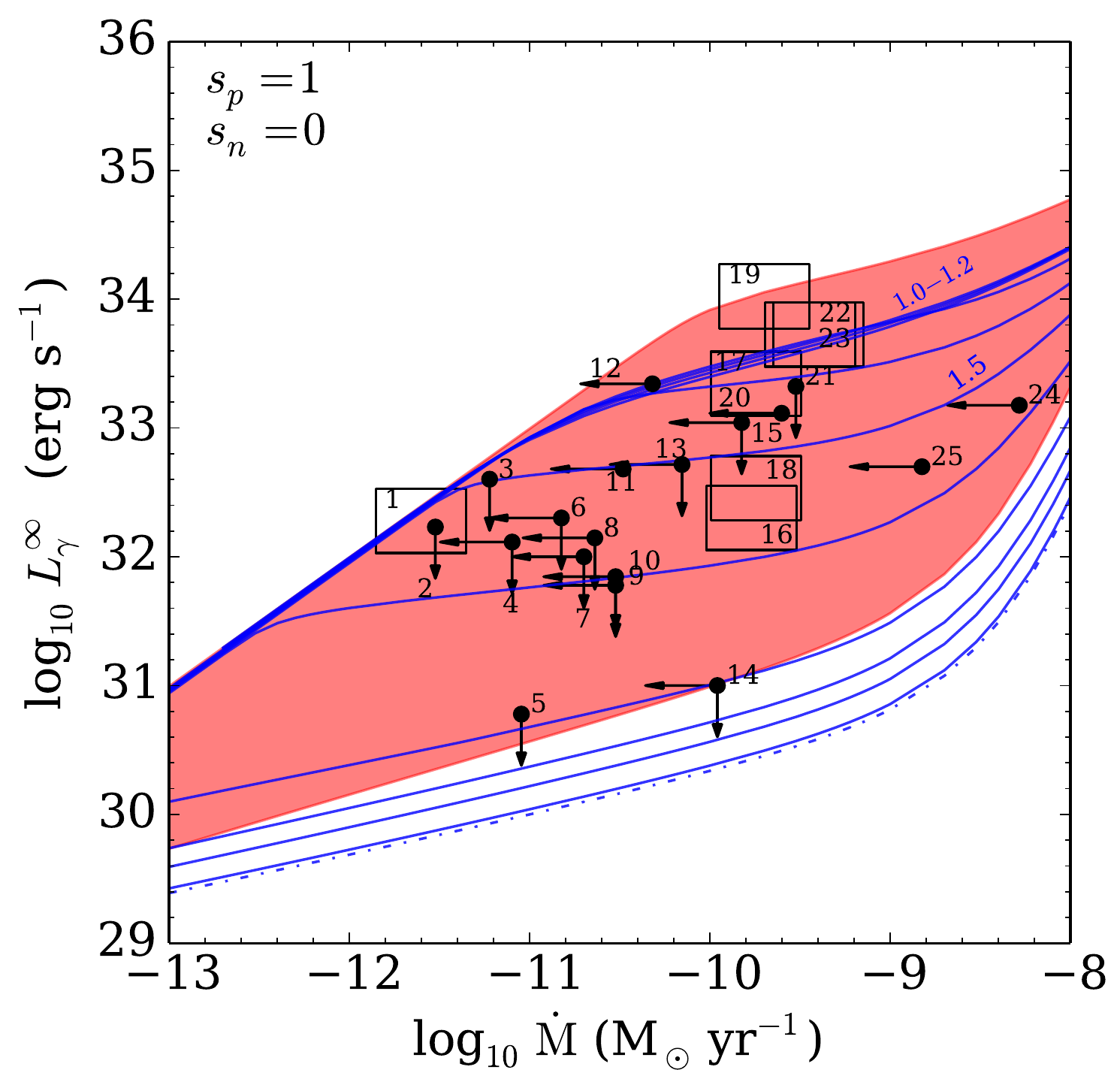}
\caption{
Same as Fig.~\ref{fig:111}
with unscaled proton superfluidity but no 3P2 neutron superfluidity:
$s_p=1$, $s_n=0$.}
\label{fig:101}
\end{figure*}

\begin{figure*}
\includegraphics[width=\columnwidth]{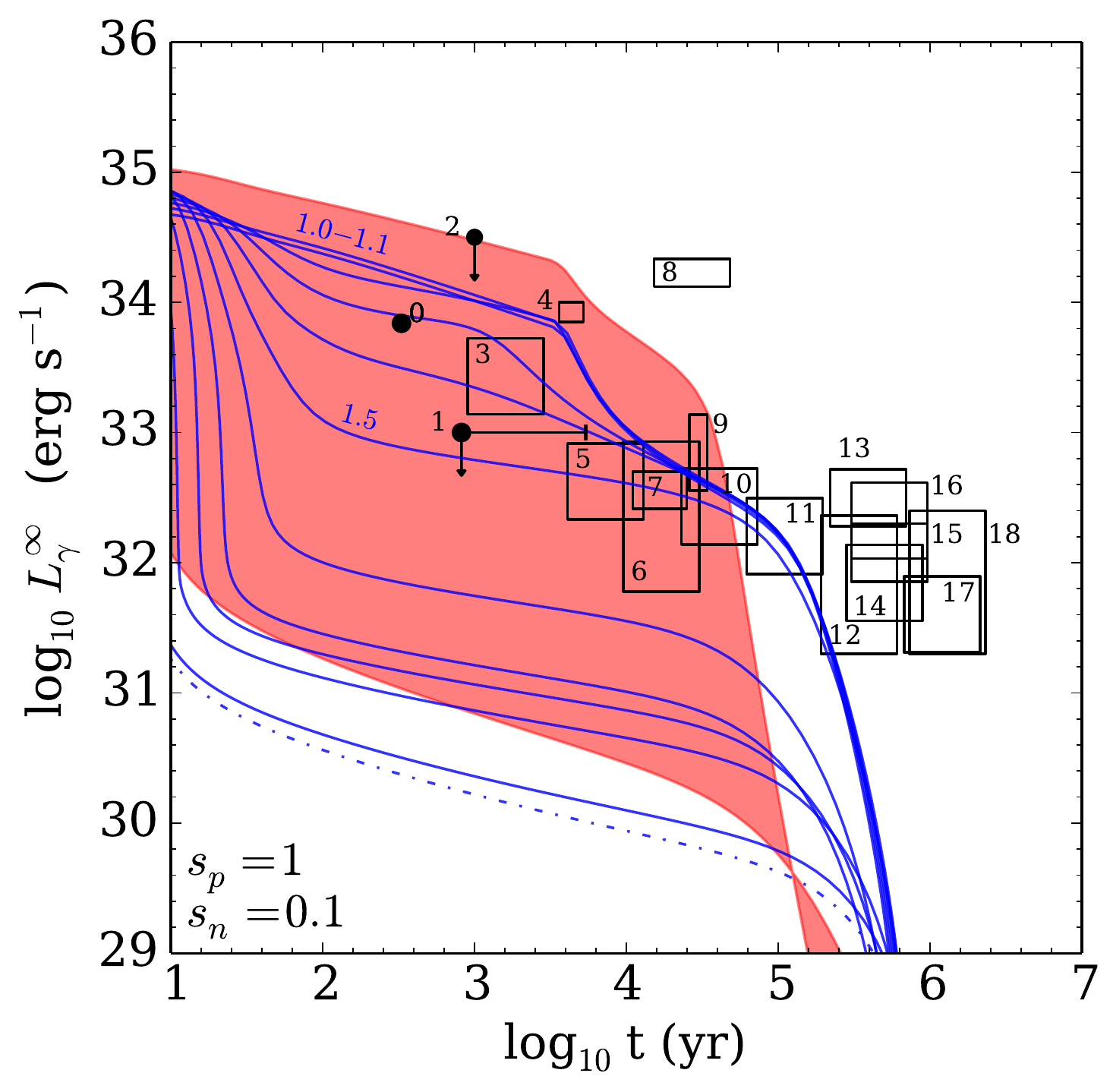}
\includegraphics[width=\columnwidth]{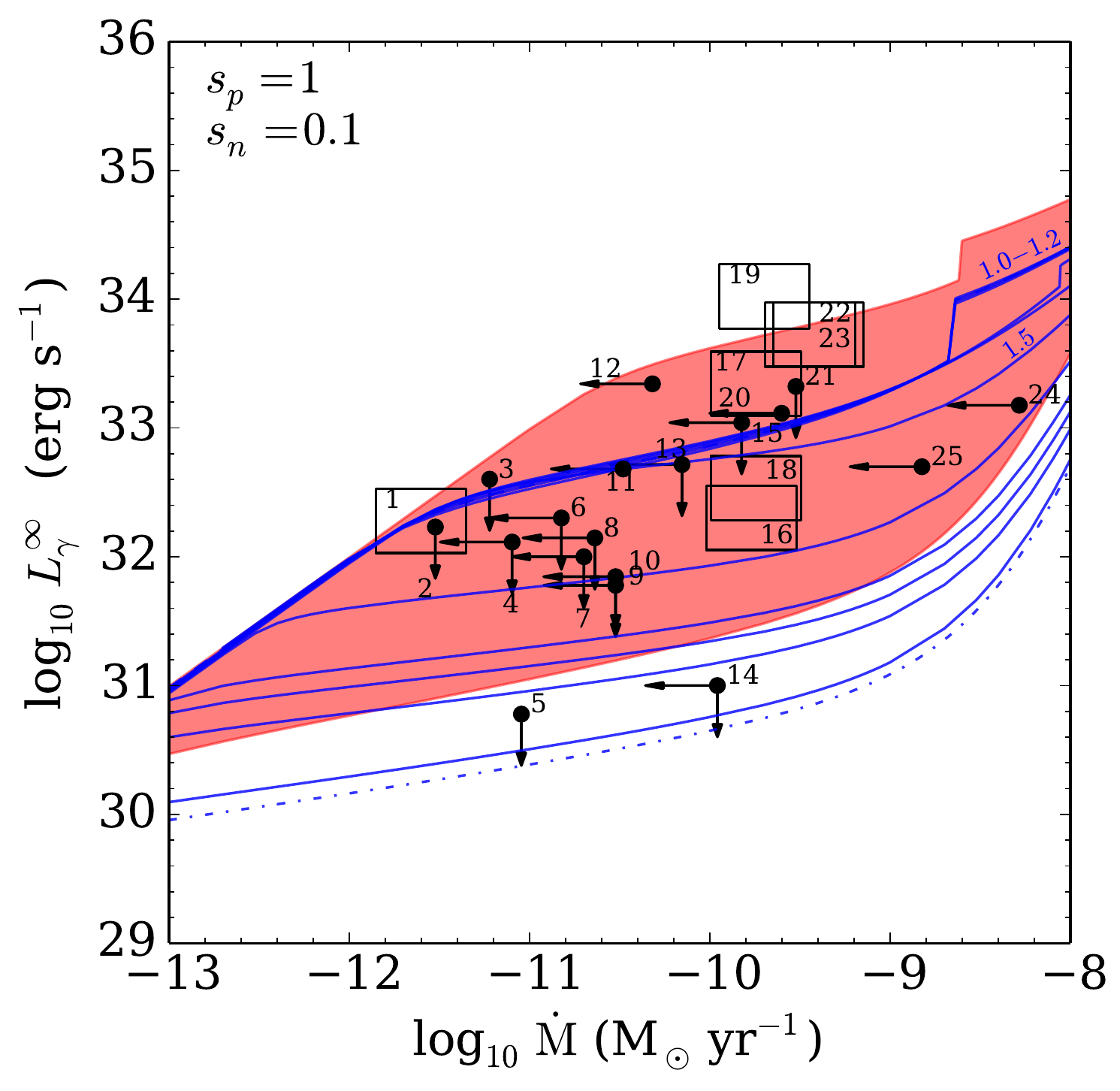}
\caption{
Same as Fig.~\ref{fig:111}
with unscaled proton superfluidity and scaled 3P2 neutron superfluidity:
$s_p=1$, $s_n=0.1$.}
\label{fig:1011}
\end{figure*}

\begin{figure*}
\includegraphics[width=\columnwidth]{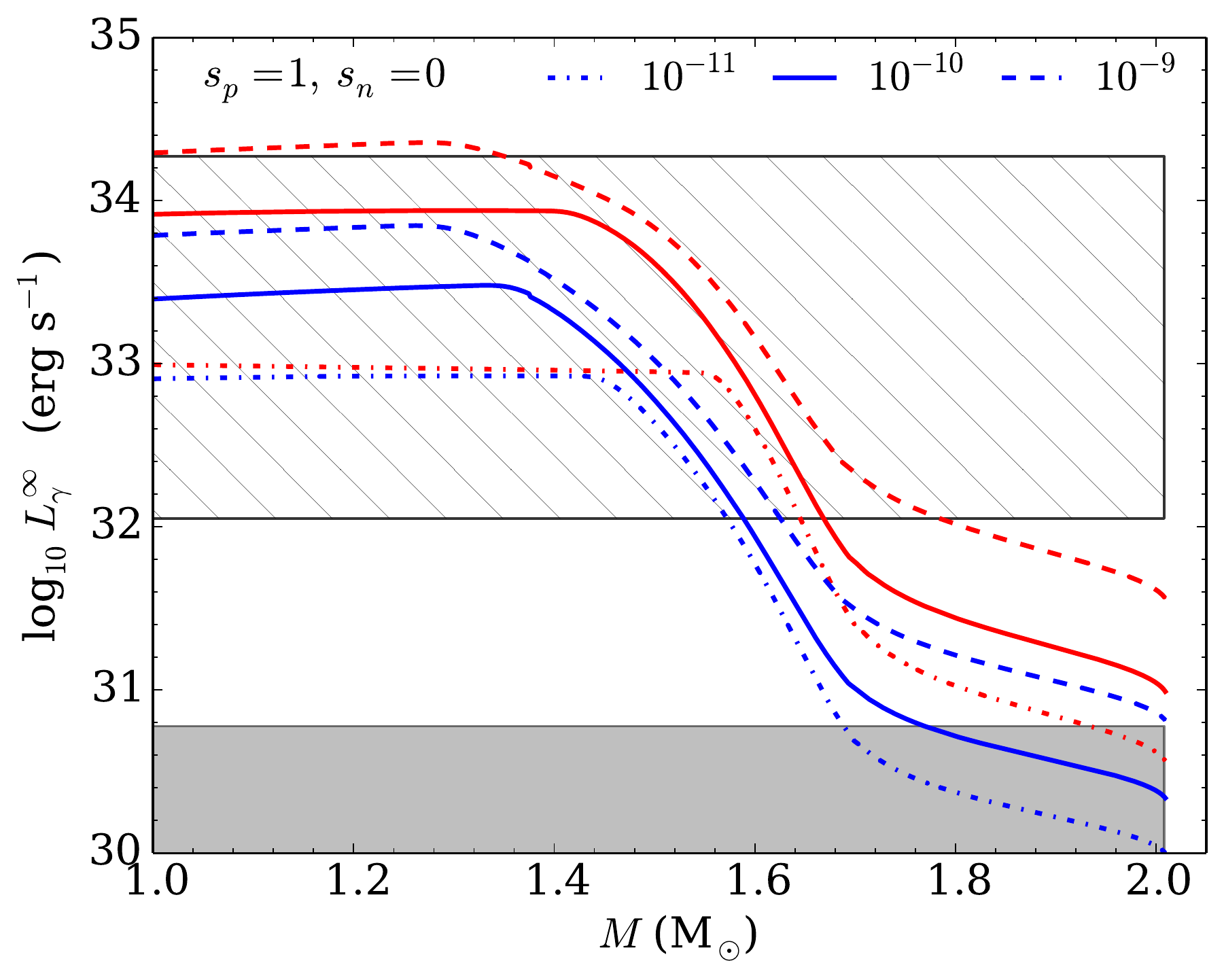}
\includegraphics[width=\columnwidth]{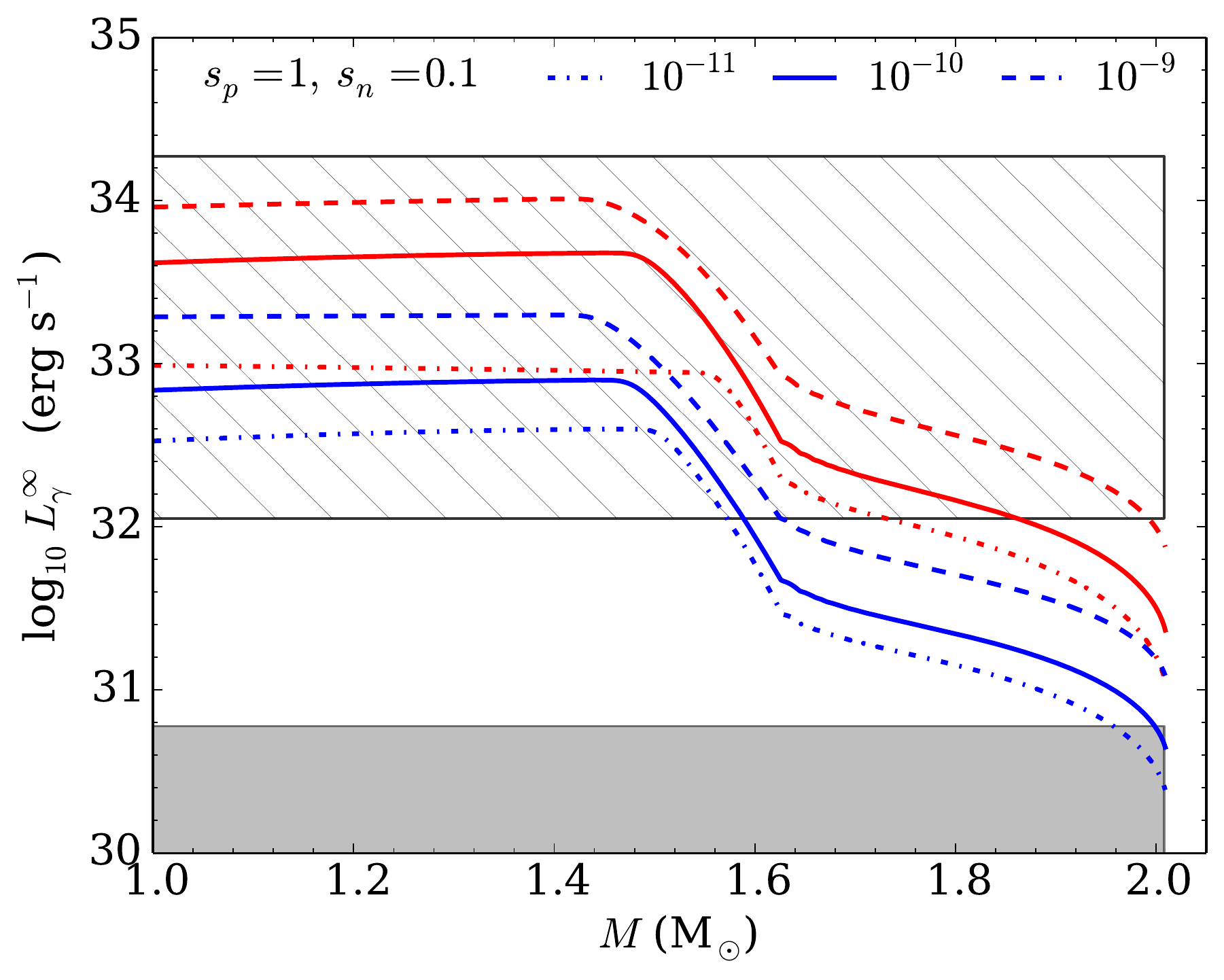}
\caption{
Same as Fig.~\ref{fig:111m}
with unscaled proton superfluidity ($s_p=1$)
and scaled 3P2 neutron superfluidity:
$s_n=0$ (left) and $s_n=0.1$ (right).}
\label{fig:101m}
\end{figure*}

We now model the thermal states of INS and qXRT using an EOS and superfluid gaps
that are consistently calculated.
Cooling and heating curves are presented in
Figs.~\ref{fig:111} and \ref{fig:1s1s1s},
without and with effective mass effects included,
$s_p=1$, $s_n=1$ and $s_p=1^*$, $s_n=1^*$, respectively,
and the luminosity-mass relation for a selection of interesting qXRT
is plotted in Fig.~\ref{fig:111m}.

First, let us have a look at the thermal states of NS with $M<\mdu$,
e.g., $M=1.0\,\ms$.
As mentioned in Section~\ref{sect:micro},
the effect of superfluidity is twofold.
On one hand the specific heat and neutrino emissivity of processes
to which the superfluid baryons contribute
are strongly reduced
when the temperature is smaller than the superfluid critical temperature.
On the other hand,
superfluidity triggers the additional n3P2 PBF process in the core,
which compensates the effects of the superfluid reduction
of the neutrino emissivity and specific heat.
It should be noted that compared to previous works not including
the anomalous contributions to the neutron PBF process,
the reduction of the PBF neutrino emissivity by a factor four
following \cite{L16} reduces the neutrino losses
and thus results in slightly larger luminosities.

All in all, in the absence of DU processes superfluidity accelerates
the cooling of an INS (see, e.g., \citealt{P04,P09})
while its temperature passes through the relevant critical temperature,
resulting at later times in lower $\lgi$ compared to the non-superfluid case,
as can be seen in Figs.~\ref{fig:111} and \ref{fig:1s1s1s}.
Similarly the $\lgi$ of qXRT with $M<\mdu$ is lower
if superfluid effects are included,
since the total neutrino losses are then larger due to the PBF processes.
For the gaps used in this work the main effect of superfluidity
on NS with $M>\mdu$ is the strong reduction of the DU process,
as can be seen in Fig.~\ref{fig:DU},
and thus a substantial reduction of the total neutrino emissivity.
Hence INS and qXRT with a mass above the DU threshold have a higher
$\lgi$ compared to the non-superfluid case,
and consequently the range of luminosities covered in the $\lgi-t$
and $\lgi-\dot{M}$ planes are much smaller.

When the effective mass effects are included,
the critical temperature of the n3P2 superfluidity is larger
at high densities $n_B\geq0.7\,$fm$^{-3}$, i.e., $M>1.7\,\ms$,
and thus the reduction of the DU process is larger for high masses,
which results in higher values $\lgi$ for both INS and qXRT
compared to the case with $s_p=1$, $s_n=1$.
However, for NS with masses $M<\mdu$,
the PBF process is more powerful,
since the leading neutrino process is the n3P2 PBF process,
and since the critical temperature is smaller for $s_p=1^*$, $s_n=1^*$.
This results in higher neutrino losses and thus lower $\lgi$
compared to the case without effective mass effects on the superfluid gap.
Hence the luminosity range of the cooling and heating curves
for a given age or accretion rate, respectively,
is smaller with the inclusion of the medium effects.

For superfluid NS,
the additional source of neutrino losses due to the PBF process
makes the transition to the photon cooling stage happen earlier
compared to non-superfluid stars,
and as a consequence these models cannot be reconciled
with most if not all the old and yet luminous INS~13-18.
In addition, the INS~4 can only marginally be explained
if the medium effects are included.
As shown in Fig.~\ref{fig:111m},
none of the superfluid models is compatible with the qXRT~5,
since the superfluid reduction of the DU process is too large.
The models are however able to fit the thermal states of all the other qXRT,
with the most luminous sources becoming consistent
when covered with an envelope composed of light elements.
The models with unscaled gaps also exhibit the property that
$\lgi$ is a smooth function of the mass,
since the DU threshold is no longer a step-like function but a smooth one,
see Fig~\ref{fig:111m}.
In other words,
the problem exhibited by non-superfluid stars,
that objects over a large range of luminosities
have the same unique mass very close to $\mdu$,
is solved by including superfluidity and the resulting smooth DU threshold.

Yet, the superfluid gaps consistent with the EOS are such
that the thermal states of superfluid NS are not consistent
with the observational data on INS
(as already pointed out in \citealt{Taranto2016})
and even more on qXRT.
Using the approach presented in this reference,
we will in the following scale the superfluid gaps consistently calculated
with the EOS
and confront superfluid NS thermal states with the observational data.

\subsection{Influence of neutron superfluidity}

Let us now examine the influence of the n3P2 superfluidity,
by scaling its critical temperature as a function of the density,
while keeping the p1S0 gap unchanged.
In the following only gaps not including the effective mass effects will be used.

First we consider the case of no superfluidity in the n3P2 channel,
$s_n=0$ and $s_p=1$.
Cooling and heating curves are presented in Fig.~\ref{fig:101} and
relations between the mass and luminosity for qXRT in Fig.~\ref{fig:101m} (left).
Compared to the case with $s_n=1$, $s_p=1$,
the absence of neutron superfluidity makes low-mass objects more luminous,
as the efficient neutron PBF process is not triggered,
and massive stars become less luminous,
since the DU process is not longer reduced for
densities $n_B\gtrsim0.7\,$fm$^{-3}$ or $M>1.7\,\ms$,
as can also be seen in Fig.~\ref{fig:DU}.
The smooth DU threshold also makes the $\lgi$ a smooth function of $M$,
a given luminosity for a fixed accretion rate corresponding to a precise mass.
The case $s_p=1$, $s_n=0$ is then consistent with all the observational data
of INS and qXRT with a smooth mass distribution,
except (albeit marginally) the hot middle-aged INS~8.
In particular the low-luminous qXRT~5 is consistent with a massive NS with a
non-accreted envelope or a NS with $M\approx\mmax$ and a fully-accreted envelope.

Second, the n3P2 gap is taken to be the one obtained without medium effects,
but scaled by a factor 0.1:
$s_n=0.1$ and $s_p=1$.
Thermal states of INS and qXRT and the luminosity for a set of qXRT
are plotted as a function of the NS mass in Figs.~\ref{fig:1011}
and \ref{fig:101m} (right), respectively.
When the temperature inside a cooling INS drops below the
neutron critical temperature, or similarly,
if the temperature inside a qXRT for which Eq.~(\ref{eqn:heatANS}) is fulfilled
is below $T_c$,
then the neutron superfluid properties affect the thermal states,
explaining the differences observed when comparing the models with
$s_n=0$ and $s_n=0.1$, together with $s_p=1$,
for large ages and small accretion rates.
Neutron superfluidity triggers the neutron PBF process,
making low-mass INS and qXRT less luminous than in the case with $s_n=0$.
Thus INS~8 and 13 to 18 can not be explained.
In order to explain these objects the neutron critical temperature
should be even smaller or neutrons should not be superfluid at all
at density $n_B\simeq(0.4-0.5)\,$fm$^{-3}$
corresponding to NS masses larger than $1\,\ms$.
In addition, for $s_n=0.1$ the DU process is always suppressed by superfluidity
(see the DU threshold in Fig.~\ref{fig:DU})
and high-mass INS and qXRT have a higher $\lgi$ than those
with no neutron superfluidity.
Consequently, the qXRT~5 can only be interpreted as being a NS with
$M\approx\mmax$ with a fully-accreted envelope.
Thanks to the smoothing of the DU threshold as observed in Fig.~\ref{fig:DU},
$\lgi(M)$ is a smooth function as in the previous case.

\begin{figure*}
\includegraphics[width=\columnwidth]{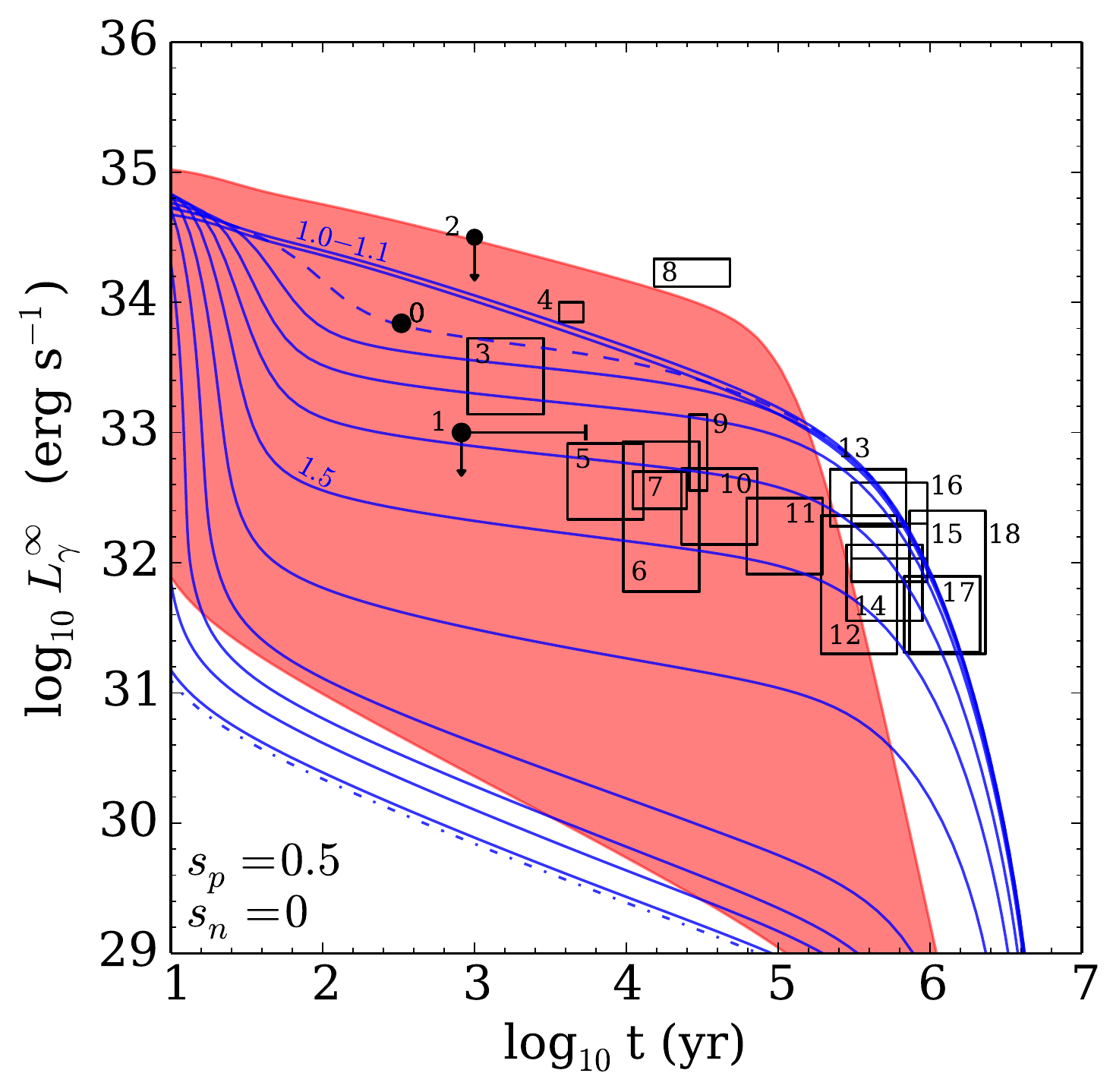}
\includegraphics[width=\columnwidth]{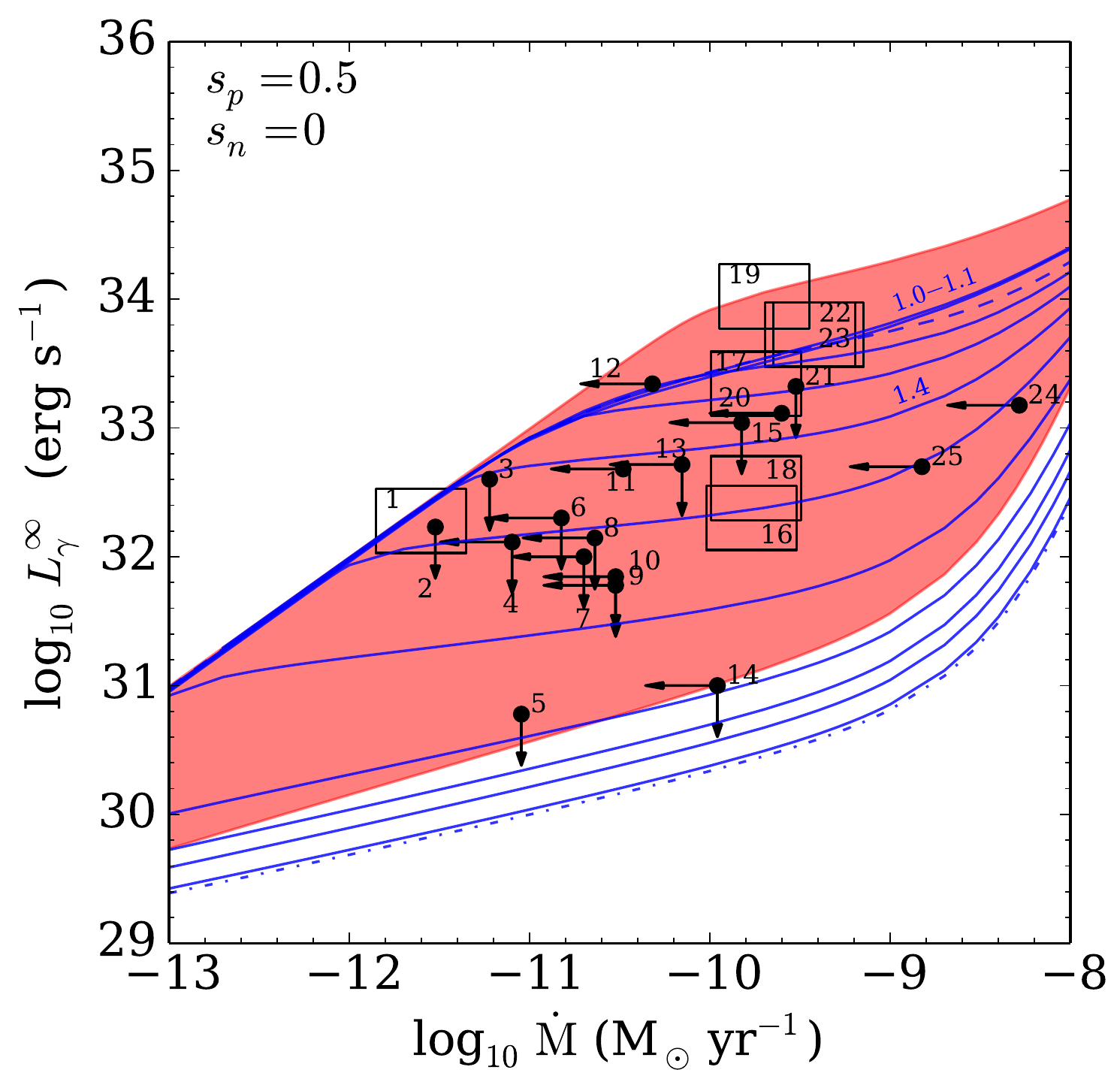}
\caption{
Same as Fig.~\ref{fig:111} for no neutron superfluidity ($s_n=0$)
and a scaled $^1$S$_0$ proton superfluidity $s_p=0.5$.}
\label{fig:1005}
\end{figure*}

\begin{figure}
\includegraphics[width=\columnwidth]{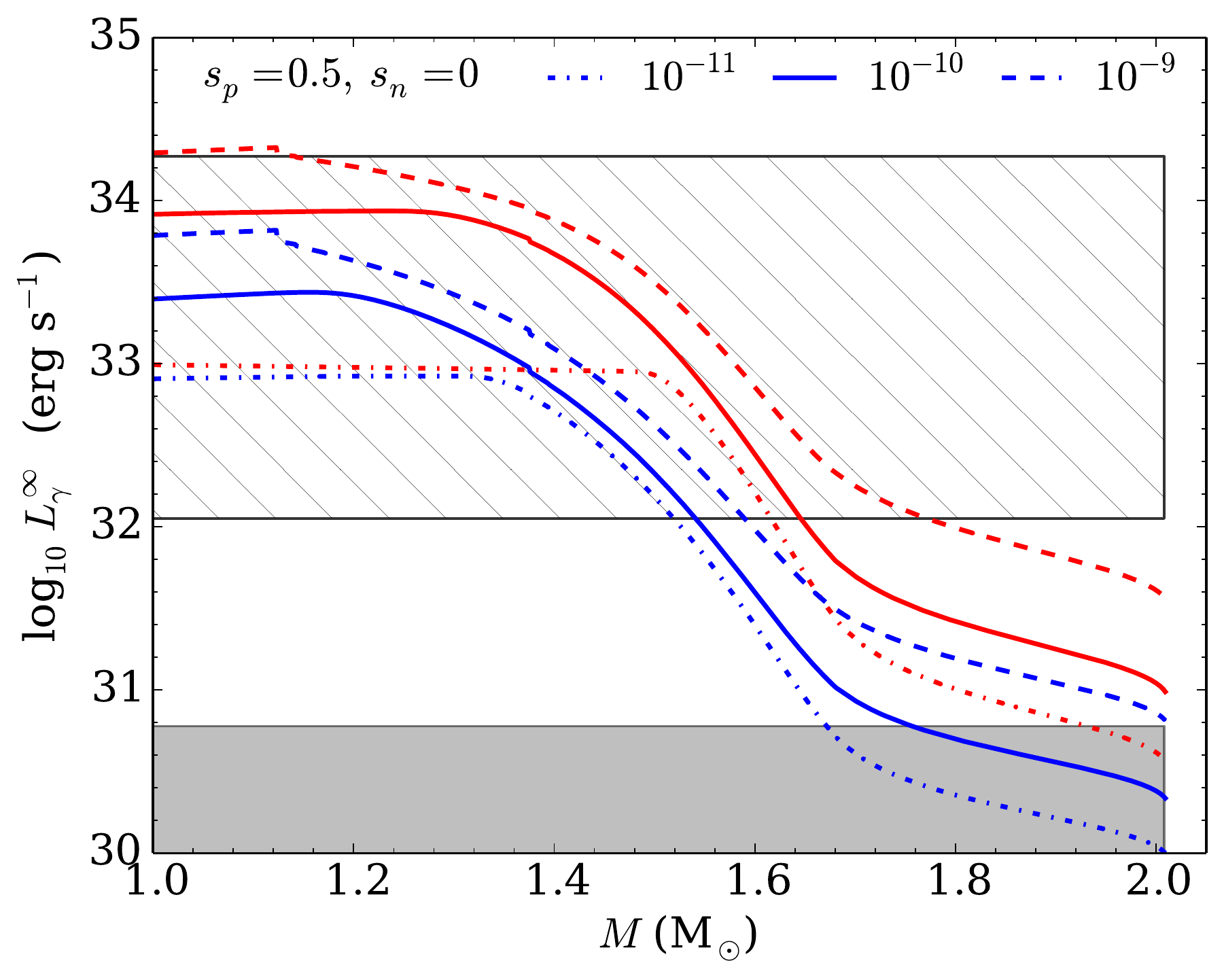}
\caption{
Same as Fig.~\ref{fig:111m} for no neutron superfluidity ($s_n=0$)
and a scaled $^1$S$_0$ proton superfluidity $s_p=0.5$.}
\label{fig:1005m}
\end{figure}

\subsection{Influence of proton superfluidity}

Let us finally investigate the influence of the proton superfluidity on the
thermal states of NS.
We do not include neutron superfluidity ($s_n=0$),
as it has been shown to better reproduce the observational data
in the previous section,
and scale the proton superfluid gap with
no effective mass effect by a factor 0.5, $s_p=0.5$.
Results are shown in Figs.~\ref{fig:1005} and \ref{fig:1005m}.

Differences with the case $s_p=1$ and $s_n=0$ are very small
and the model is consistent with all the observational data
(marginally with the INS~8).
The main effect of the reduced proton superfluidity is that the DU process
is suppressed over a smaller range of densities than in the case
with unscaled proton gap as shown in Fig.~\ref{fig:DU}.
This consequently makes the NS in which the DU process is turned on
less luminous for a given mass
(see, e.g., the cooling curve of $1.5\,\ms$ NS
in Figs.~\ref{fig:101} and \ref{fig:1005}).
Hence lower $\lgi$ can be reached at lower mass
when the proton superfluidity is reduced.

We also computed thermal states of NS for a model with a proton superfluid
gap scaled by a factor 0.5 and a neutron one taken unscaled,
$s_p=0.5$ and $s_n=1$.
Results are not shown, but thermal states of old INS with a high $\lgi$,
objects~13 to 18,
and of the low luminous qXRT~5 cannot be explained then
because of the strong neutron PBF process and the reduction
of the DU process at high density, respectively.

\section{Discussion and conclusions}

In conclusion, for the BHF EOS with the DU process operating in all currently observed NS
consistency with the thermal states of INS and NS in qXRT can be obtained if:
(a) the neutrons are not superfluid or they are
superfluid over a reduced range of densities corresponding to medium NS masses,
i.e., in low-mass NS with $M<\mdu$ and early enough in massive stars;
(b) the protons are superfluid
(the precise magnitude of the gap being of little importance),
but not in massive stars.
No or little neutron superfluidity in low-mass stars together with proton
superfluidity is required to slow down the cooling of middle-aged NS and thus
to explain the thermal state of INS~8, XMMU J1731$-$347,
as already pointed out in \cite{XMMUa,XMMUb},
and of INS~13 to 18 and qXRT~19.
Otherwise the triggering of the neutron PBF process or the
non-superfluid reduction of the slow neutrino processes
make the thermal states of low-mass NS inconsistent with these objects.
In any case INS~8 and qXRT~19 (SAX J1750.8$-$2900)
require a fully-accreted envelope,
while INS~17 (PSR J2043+2740) and 18 (RX J0720.4$-$3125) a non-accreted one.

Both neutron and proton superfluidity have to be reduced early enough
in massive stars in order to ensure that the DU process is fully operating
so that the low luminosity of qXRT~5 (SAX J1808$-$3658) can be explained as
being a NS with a high mass.
The medium mass $M\lesssim 1.7\ms$ derived for this object in \cite{SAXb}
suggests that: a) the DU process
is already operating in low-mass NS, as in our model,
but reduced in such objects by baryon superfluidity,
in order to reproduce the thermal state of XMMU J1731$-$347;
b) that the baryon superfluidity is small in medium-mass NS $M\sim 1.4\ms$
so that SAX J1808$-$3658 is consistently modelled.

All in all, as also stated in (\citealt{LH07,BY15a,HS17}),
qXRT sources require going beyond the so-called minimal cooling paradigm
\citep{Gus04,P04,P09},
where all the necessary microphysical ingredients are included
(in particular superfluidity),
but fast cooling processes such as the DU process are not.
Indeed even in case of strongly superfluid neutrons and protons,
the neutrino losses due to the PBF processes are not large enough
to explain the thermal state of qXRT~5,
even more when taking into account the anomalous contributions
to the PBF processes.
Finally strong or moderate superfluidity for densities
larger than the DU threshold appears required,
so that the DU threshold is not a step-like function,
but that the mass distribution of qXRT is smooth,
ensuring that most NS do not have a mass close to $\mdu$.

It should be mentioned that the observational data we use in this work
is subject to systematic and statistical errors.
Two objects are particularly constraining for our model:
XMMU J1731$-$347 and SAX J1808$-$3658.
Thus future observations may affect our conclusions regarding
the density profiles of the neutron and proton superfluidity in the core.
Yet we argue that the broadening of the DU threshold can be obtained
by taking into account the baryon superfluid properties
in the modeling of the thermal states of INS and qXRT,
as was done in the present work.
The so-called nuclear-medium modification of the thermal properties
\citep{MMa,MMb} can lead to a similar smoothing of the mass distribution
[see also the discussion in \cite{BY15a,BY15b}]
and will be the topic of a future work.

There are equally successful cooling calculations
based on very different nuclear EOSs and pairing gaps,
but what we would like to point out is that even making some extreme
(but theoretically justified and consistent)
assumptions about the symmetry energy and cooling processes
(DU active in all NS),
our model cannot be falsified by the current data on NS cooling.
Furthermore and more generally,
in our work we used a minimal model of NS,
assuming purely nucleon cores.
We were able to obtain a reasonable agreement with existing observations.
So for the time being,
and in the context of observations of cooling INS and qXRT,
there seems to be no urgent need to go beyond the minimal NS model,
and to consider exotic NS cores (hyperons, quarks, boson condensates).
This does not mean that the minimal model represents reality.
In particular, if future theoretical calculations firmly demonstrate
significant neutron pairing at high density,
one might be required to abandon it.

\section*{Acknowledgements}

This work was partially supported by the Polish NCN research grant OPUS6
no.~2013/11/B/ST9/04528.
We also acknowledge partial support from the NewCompStar COST Action MP1304.




%

\bsp	
\label{lastpage}
\end{document}